\documentclass[journal]{IEEEtran}
\usepackage{amsmath,amsfonts,amsthm,amssymb,mathrsfs}
\usepackage{algorithmic}
\usepackage{array}
\usepackage[caption=false,font=normalsize,labelfont=sf,textfont=sf]{subfig}
\usepackage{textcomp}
\usepackage{stfloats}
\usepackage{url}
\usepackage{verbatim}
\usepackage{graphicx}
\usepackage{bbding}
\usepackage{multirow}
\usepackage{cite, bm, bbm, xcolor}
\usepackage{refcount}
\usepackage{flafter}
\usepackage{makecell}
\usepackage{diagbox}
\usepackage{booktabs}
\usepackage{hyperref}
\newtheoremstyle{custom_theoremstyle}{\topsep}{\topsep}{\upshape}{1em}{\itshape}{:}{5pt plus 1pt minus 1pt}{}
\theoremstyle{custom_theoremstyle}

\newtheorem{lemma}{Lemma}
\newtheorem{remark}{Remark}
\theoremstyle{definition}
\newtheorem{definition}{Definition}
\usepackage{algorithm}

\usepackage{subcaption}
\usepackage{ulem}

\usepackage[T1]{fontenc}
\providecommand{\url}[1]{#1}
\allowdisplaybreaks[4]

\begin{document}
\title{Joint Transmit and Pinching Beamforming for Pinching Antenna System (PASS): Optimization-Based or Learning-Based?}

\author{
Xiaoxia Xu, \textit{Member, IEEE},  Xidong Mu, \textit{Member, IEEE},
Yuanwei Liu, \textit{Fellow, IEEE}, and Arumugam Nallanathan, \textit{Fellow, IEEE}
\thanks{This work was jointly supported
by the Engineering and Physical Sciences Research Council (EPSRC) under Grant EP/W004100/1, Grant EP/W034786/1, and Grant EP/Y037243/1. 
Part of this work has been presented at IEEE/CIC International Conference on Communications in China (ICCC) in 2025 \cite{PASS_ICCC}.
\textit{(Corresponding Author: Arumugam Nallanathan)}}
\thanks{X. Xu is with the School of Electronic Engineering and Computer Science, Queen Mary University of
London, London E1 4NS, U.K. (e-mail: x.xiaoxia@qmul.ac.uk).}
\thanks{X. Mu is with the Centre for Wireless Innovation (CWI), Queen's University Belfast, Belfast, BT3 9DT, U.K. (e-mail: x.mu@qub.ac.uk)}
\thanks{Y. Liu is with the Department of Electrical and Electronic Engineering, the University of Hong Kong, Hong Kong, and also with Faculty of Applied
Sciences, Macao Polytechnic University (e-mail: yuanwei@hku.hk).}
\thanks{A. Nallanathan is with the School of Electronic Engineering and Computer Science, Queen Mary University of London, London 
and also with the Department of Electronic Engineering, Kyung Hee University, 
Yongin-si, Gyeonggi-do 17104, Korea (e-mail: a.nallanathan@qmul.ac.uk).}
}

\maketitle

\vspace{-1.5em}
\begin{abstract}  
    A novel pinching antenna system (PASS)-enabled downlink multi-user multiple-input single-output (MISO) framework is proposed. 
    PASS consists of multiple waveguides spanning over thousands of wavelength, which equip numerous low-cost dielectric particles, named pinching antennas (PAs), to radiate signals into free space. 
    The positions of PAs can be reconfigured to change both the large-scale path losses and phases of signals, 
    thus facilitating the novel \textit{pinching beamforming} design.
    A sum rate maximization problem is formulated, which jointly optimizes the transmit and pinching beamforming to adaptively achieve constructive signal enhancement and destructive interference mitigation.  
    To solve this highly coupled and nonconvex problem, both optimization-based and learning-based methods are proposed. 
    1) For the optimization-based method, a majorization-minimization and penalty dual decomposition (MM-PDD) algorithm is developed, 
    which handles the nonconvex complex exponential component using a Lipschitz surrogate function and then invokes PDD for problem decoupling. 
    2) For the learning-based method, a novel Karush-Kuhn-Tucker (KKT)-guided dual learning (KDL) approach is proposed, 
    which enables KKT solutions to be reconstructed in a data-driven manner by learning dual variables. 
    Following this idea, a KDL-Transformer algorithm is developed, which captures both inter-PA/inter-user dependencies and  channel-state-information (CSI)-beamforming dependencies by attention mechanisms. 
    Simulation results demonstrate that: i) The proposed PASS framework significantly outperforms conventional massive multiple input multiple output (MIMO) system even with a few PAs. 
    ii) The proposed KDL-Transformer can improve over $20\%$ system performance than MM-PDD algorithm, 
    while achieving a millisecond-level response on modern GPUs. 
\end{abstract}
\begin{IEEEkeywords}
    Beamforming, deep learning, flexible antennas, pinching antenna, pinching-antenna system (PASS).
\end{IEEEkeywords}

\section{Introduction}

Shannon's theorem establishes the capacity upper bound of wireless communication systems based on the signal-to-noise ratio (SNR)~\cite{Shannon}, which dramatically depends on channel qualities of mobile users.
To break through the limitation of fixed-channel assumptions and make  radio environment customizable, 
recent advances in sixth-generation (6G) wireless networks have been devoted  
to developing several new forms of flexible-antenna techniques, such as reconfigurable intelligent surfaces (RISs)~\cite{RIS,STARS}, 
fluid antennas~\cite{FluidAntenna}, and movable antennas~\cite{MovableAntenna}.
Specifically, RISs equip a large amount of low-cost passive reflective/transmissive meta-material elements to reconfigure the communication environment 
by altering the phase shifts and amplitudes of incident signals~\cite{RIS,STARS}. 
More recently, fluid antenna and movable antenna systems have been proposed~\cite{FluidAntenna,MovableAntenna}, 
which enable the transceivers to dynamically adjust antenna positions and their electromagnetic properties, 
thus creating favorable spatial channels for multi-input multi-output (MIMO) communications.

Existing flexible-antenna technologies are effective in shaping the scattering environment to improve channel conditions.
However, since conventional MIMO transceivers typically operate within a limited spatial range, their ability to influence large-scale fading remains restricted.
Recently, the pinching antenna system (PASS) has emerged as a novel flexible-antenna technique~\cite{PAr_Ding,PAr_Liu}, which enables direct manipulation of path loss. 
PASS is built on a rod-like transmission line, known as a dielectric waveguide, that can span from a few meters to tens of meters, or even up to thousands of times 
longer than the wavelength~\cite{PA_DOCOMO}.
The dielectric waveguide acts as a leaky-wave antenna, along which multiple radiation points, termed pinching antennas (PAs), can be flexibly pinched and released at arbitrary locations.
These PAs are implemented by small dielectric particles non-contact to the waveguides. 
Different from conventional leaky-wave systems with fixed radiating elements, PASS enables dynamic and reconfigurable activation and placement of these PAs 
in response to varying user and channel conditions. 
This unique design of PASS leads to several advantages~\cite{PAr_Liu}: 
\textit{1) Path-loss reconfiguration and LoS link construction}: PASS can reconfigure the large-scale path loss 
to deliver wireless services to the last-meter range, 
and maintain stable line-of-sight (LoS) links to avoid blockage in high-frequency communications. 
\textit{2) Flexible beamforming}: The deployment of PAs can transform the entire electromagnetic propagation environment, 
thus achieving flexible beam focusing/steering effects even with a few PAs. 
\textit{3) Scalability}: PAs can be easily activated, shifted, and released from the dielectric waveguides to accommodate various traffic demands.  

Owing to the above appealing features, PASS recently becomes a focal point in both academia and industry. 
The first prototype of PASS has been developed by DOCOMO in 2021, and its effectiveness at 60-GHz millimeter-wave (mmWave) band has been tested~\cite{PA_DOCOMO}.
The authors of~\cite{PAr_Ding} conceived both single-waveguide architecture and multi-waveguide architecture for PASS. 
For multi-waveguide architecture, a single PA is activated on each waveguide, and the performance upper bound of the corresponding two-user multi-input single-output (MISO) interference channel have been derived.  
Furthermore, the authors of~\cite{Rate_DL_PAr_SU} considered the downlink sum rate maximization problem of a single-user PASS scenario.
The locations of multiple PAs are optimized by a two-stage algorithms, where the first stage determines PAs' locations by minimizing the large-scale path loss, 
and the second stage refines PAs' locations to ensure constructive interference. 
In~\cite{Optimal_spacing}, the authors demonstrated the existence of an optimal inter-antenna spacing to maximize the array gains of PASS. 
Moreover, the authors of~\cite{Discrete_matching} considered a discrete single-waveguide PASS architecture, 
where the activated PAs attached at predefined locations were selected to activate by one-to-one matching. 

Note that the aforementioned literatures mainly focused on either single-waveguide designs or multiple-waveguide designs where only single PA is activated at each waveguide.
Up to date, how to develop a general PASS architecture for multi-user MISO communications is still less understood. 
The remaining open challenges are two-fold: 
\begin{itemize}
    \item The deployment of each PA impacts both the large-scale path loss and the phases of signals, which leads to the novel concept of \textit{pinching beamforming}.
    Despite providing additional degree-of-freedom (DoF), efficient pinching beamforming is required to 
    simultaneously alter large-scale fading and signal phases for both path loss reduction and multi-user interference mitigation, 
    which is challenging even in a LoS-dominant environment. 
    \item Activating multiple PAs along waveguides leads to constructive signal enhancement and destructive interference mitigation of radio waves. 
    Hence, the cooperative deployment of PAs is crucial for multi-user communications in PASS. 
    However, how to achieve fast cooperative deployment of multiple PAs remains to be explored, especially when the number of PAs increases. 
\end{itemize}

To address above issues, this paper proposes a novel PASS-enabled downlink multi-user MISO framework. %, which supports both pinching beamforming and transmit beamforming. 
A sum rate maximization problem is formulated by jointly optimizing the newly introduced pinching beamforming and the conventional transmit beamforming. 
The resulting optimization problem is highly coupled and involves nonconvex complex exponential components. 
To tackle this problem, this paper proposes both optimization-based and learning-based methods. 
\textit{First}, by reformulating the coupled optimization problem using the weighted minimum mean square error (WMMSE), 
we develop a majorization-minimization and penalty dual decomposition (MM-PDD) algorithm, which is guaranteed to find the stationary solution. 
Specifically, the Lipschitz gradient surrogate function is constructed for convex relaxation of complex exponential components.   
Then, the coupled optimization is decomposed by PDD theory and iteratively solved by block coordinate descent (BCD). 
\textit{Secondly}, to achieve fast beamforming with a millisecond-level response, we propose a novel optimization theory-guided learning method, 
termed \textit{Karush-Kuhn-Tucker (KKT)-guided dual learning (KDL)}. 
The key idea is to enable machine learning (ML) to directly reconstruct KKT solutions by learning only a few dual variables in a data-driven manner. 
Hence, it inherits the white-box KKT solution structure, but avoids inefficient alternating block descents required by iterative optimization algorithms.  
We implement KDL by exploiting the powerful large language model (LLM) techniques, and comes up with a KDL-Transformer algorithm. 
By casting the channel-state-information-(CSI)-to-beamforming mapping as a sequence-to-sequence learning task, 
both inter-PA/inter-user and CSI-beamforming dependencies over structured input and output sequences can be jointly captured. 
The proposed KDL-Transformer achieves substantial performance gains over both black-box learning and mathematical optimization methods.

The main contributions can be summarized as follows.
\begin{itemize}
    \item We propose a novel PASS-enabled downlink multi-user MISO framework, 
    which enables the additional pinching beamforming by configuring PAs' locations. 
    A joint transmit and pinching beamforming optimization problem is formulated to maximize the system sum rate achieved by PASS. 
    We propose both optimization-based and learning-based methods to tackle this problem.   
    \item For the optimization-based method, we propose an MM-PDD algorithm that efficiently addresses the highly nonconvex and coupled problem with strong theoretical guarantees. 
    By reformulating the original problem into a more tractable WMMSE problem, 
    we derive Lipschitz gradient surrogate function based on MM to handle the nonconvex complex exponential component, 
    and then decouple the problem based on PDD. We mathematically prove that the proposed algorithm converges to stationary solutions. 
    \item For the learning-based method, we propose a novel KDL approach. 
    KDL learns dual variables to reconstruct KKT solutions in a low-complexity and data-driven manner, 
    instead of inferring all primal variables by black-box models.
    To unleash the potential based on LLM techniques, we develop a KDL-Transformer, which can 
    capture both inter-PA/inter-user dependencies and CSI-beamforming dependencies to find high-quality solutions. 
    \item We provide numerical results to demonstrate the effectiveness of the proposed framework and algorithms, 
    which demonstrates that: 1) The proposed PASS architecture outperforms conventional hybrid beamforming architecture 
    even only a few PAs are activated. 2) While black-box learning algorithms are stuck at inferior points, the proposed KDL-Transformer increases 
    over $20\%$ system performance than MM-PDD algorithm, while achieving millisecond-level fast inference on modern GPU devices. 
\end{itemize}

The rest of this paper is organized as follows. 
Section II presents the proposed PASS framework for downlink multi-user MISO communications and formulates the sum rate maximization problem. 
Section III develops an MM-PDD algorithm for joint digital and pinching beamforming, and Section IV proposes the KKT-guided dual learning method and develops a KDL-Transformer algorithm. 
Section V provides numerical results to verify the efficiency of the proposed framework and algorithms. 
Finally, Section VI concludes the paper.

\textit{Notations}: 
Lowercase letters (e.g., $x$) denote scalars, bold lowercase letters (e.g., $\mathbf{x}$) denote vectors, 
and bold uppercase letters (e.g., $\mathbf{X}$) denote matrices.
$|x|$ denotes the absolute value of a real number or the modulus of a complex number. 
$\mathrm{Re}\left\{x\right\}$ and $\mathrm{Im}\left\{x\right\}$ denote the real and image parts of $x$, and $x^{H}$ is the complex conjugate number of $x$.  
$\mathbf{1}_{N\times 1}$ denotes an $N$-dimension all-ones vector.
$\mathbf{X}^{\top}$ and $\mathbf{X}^{H}$  denote the transpose and the Hermitian matrix. 
$\left\Vert\mathbf{x}\right\Vert$ is the vector Euclidean norm, 
$\left\Vert\mathbf{X}\right\Vert_{\infty}=\max_{i,j}\left|x_{i,j}\right|$, 
and $\left\Vert\mathbf{X}\right\Vert$ is the matrix Frobenius norm.

\section{System Model and Problem Formulation} 

\begin{figure}[!t]
    \centering
    \includegraphics[width=0.49\textwidth]{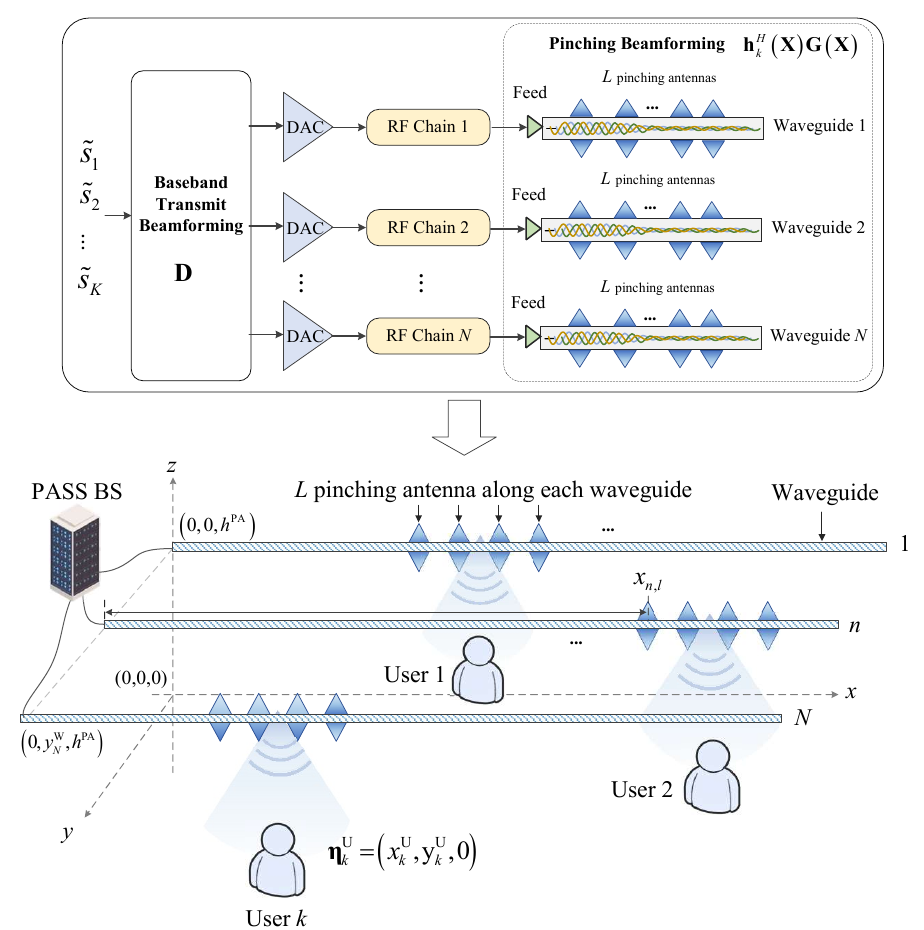}
    \caption{PASS-enabled downlink multi-user MIMO framework.}\label{fig_pass}
\end{figure}

In this paper, we propose a PASS-enabled downlink multi-user MISO framework,  
as shown in Fig. \ref{fig_pass}. The BS is equipped with $N$ waveguides and $M$ PAs to serve $K$ single-antenna users.  
Each waveguide stretches thousands of wavelengths. 
There are $L$ PAs flexibly pinched along each waveguide, and the total number of PAs is $M = L \times N$.
By activating PAs at different points along the waveguides, both the phases of incident signals and the large-scale fading can be altered. 
Let $\mathcal{N}$ and $\mathcal{M} = \mathcal{L}_{1} \cup \mathcal{L}_{2} \cup \dots \mathcal{L}_{N}$ 
denote the sets of all the waveguides and PAs, respectively, 
where $\mathcal{L}_{n} = \left\{a_{n,1}, a_{n,2}, \dots, a_{n,L} \right\}$ is the collection of PAs at waveguide $n$. 
To support spatial multiplexing,  
each waveguide is connected to a dedicated radio frequency (RF) chain, 
which converts the signal multiplexed at the baseband and feeds it into the waveguide.   
To ensure multiplexing gains, we assume $N=K$. 
Both waveguides and PAs are installed at a fixed height of $h^{\mathrm{PA}}$, 
and the resulting PASS spans across a rectangular area with a size of $S_{\mathrm{x}} \times S_{\mathrm{y}}$ $\text{m}^2$. % whose center is denoted by $\bm{\eta}_{\mathrm{c}} = \left[S_{\mathrm{x}}/2, S_{\mathrm{y}}/2, 0\right]$. 
The three-dimension Cartesian coordinate of the feed point for the $n$-th waveguide 
is given by $\bm{\eta}_{n}^{\mathrm{W}} = \left[0, y_{n}^{\mathrm{W}},  h^{\mathrm{PA}}\right]^{\top}$, 
where $y_{n}^{\mathrm{W}}$ is the $y$-axis coordinate of this waveguide. 
The location of each PA $a_{n,l}$ along the $n$-th waveguide can be defined as 
$\bm{\eta}_{n,l}\left(x_{n,l}\right) = \left[ x_{n,l}, y_{n,l},  h^{\mathrm{PA}}\right]^{\top}$, 
where $x_{n,l}$ is the adjustable pinched location of PA $a_{n,l}$ along $x$-axis, 
and $y_{n,l}$ is the fixed coordinate over $y$-axis that depends on the connected waveguide, i.e., $y_{n,l} = y_{n}^{\mathrm{W}}$.
Let $\mathbf{x}_{n}=\left[x_{n,1}, x_{n,2}, \dots, x_{n,L}\right]^{\top} \in \mathbb{R}^{L \times 1}$ stack 
the $x$-axis locations of PAs over the $n$-the waveguide,  
which satisfy $0 \leqslant x_{n,1} < x_{n,2} < \dots < x_{n,L} \leqslant S_{\mathrm{x}}$, $\forall n \in \mathcal{N}$. 
$\mathbf{X}=\left[\mathbf{x}_{1},\mathbf{x}_{2},\dots,\mathbf{x}_{N}\right]\in\mathbb{R}^{N\times L}$ stacks the locations of all PAs. 
Furthermore, the location of  user $k$ is denoted by $\bm{\eta}_{k}^{\mathrm{U}} = \left[x_{k}^{\mathrm{U}}, y_{k}^{\mathrm{U}}, 0\right]^{\top}$, $\forall k \in \mathcal{K}$.

\subsection{Signal Radiation Model}
Each waveguide needs to be fed with the same signal. 
This signal is multiplexed at the baseband through the conventional transmit beamforming $\mathbf{D}$,  
and then converted by the RF chain and fed into the waveguide for radiation. 
As a result, the transmitted signal of user $k$ radiated by the PASS to the free space is given by
\begin{equation}
    \mathbf{s}_{k}^{\mathrm{TX}}
    = \mathbf{G}\left(\mathbf{X}\right)\mathbf{d}_{k}\widetilde{s}_{k} \in \mathbb{C}^{M\times 1},
\end{equation}
where $\mathbf{G}\left(\mathbf{X}\right)\in\mathbb{C}^{M\times N}$ denotes the path response when the signals are propagated from the feed point of each waveguide to specific PAs, 
$\mathbf{d}_{k} \in \mathbb{C}^{N\times 1}$ denotes the conventional transmit beamforming vector for user $k$, 
and $\widetilde{s}_{k}$ is the data signal satisfying $\mathbb{E}\left[\widetilde{s}_{k}^{H} \widetilde{s}_{k}\right] = 1$.  
As each waveguide is connected to a subset of PAs, 
$\mathbf{G}\left(\mathbf{X}\right)$ is a block-diagonal matrix, given by 
\begin{equation}
    \begin{split}
        \mathbf{G}\left(\mathbf{X}\right)&= 
        \mathrm{blkdiag} \left(\mathbf{g}_{1}\left(\mathbf{x}_{1}\right),\mathbf{g}_{1}\left(\mathbf{x}_{2}\right), \dots, \mathbf{g}_{N}\left(\mathbf{x}_{N}\right)\right)
        \\
        &=
        \left[\begin{array}{cccc}
            \mathbf{g}_{1}\left(\mathbf{x}_{1}\right) & \mathbf{0} & \dots & \mathbf{0}\\
            \mathbf{0} & \mathbf{g}_{2}\left(\mathbf{x}_{2}\right) & \dots & \mathbf{0}\\
            \vdots & \vdots & \ddots & \vdots\\
            \mathbf{0} & \mathbf{0} & \dots & \mathbf{g}_{N}\left(\mathbf{x}_{N}\right)
            \end{array}\right],
    \end{split}
\end{equation}
where $\mathbf{g}_{n}\left(\mathbf{x}_{n}\right) \in \mathbb{C}^{L\times 1}$ is the 
response vector from the feed point of waveguide $n$ to the associated subset of PAs in $\mathcal{L}_{n}$. %(along $x$-axis). 
The $l$-th element of $\mathbf{g}_{n}\left(\mathbf{x}_{n}\right)$ is given by
\begin{equation}
    g_{n,l}\!\left(x_{n,l}\right) \!\!=\!\!
    \frac{1}{\sqrt{L}}e^{-i\frac{2\pi}{\lambda_{\mathrm{W}}}\left\Vert \bm{\eta}_{n,l}\left(x_{n,l}\right)-\bm{\eta}_{n}^{\mathrm{W}}\right\Vert}
    \!\!=\!\!\frac{1}{\sqrt{L}}e^{-i\kappa n_{\mathrm{eff}}x_{n,l}},
\end{equation}
where $\lambda_{\mathrm{W}}=\tfrac{\lambda_{f}}{n_{\mathrm{eff}}}$ is the guided wavelength with $n_{\mathrm{eff}}$ being the effective refractive index of the dielectric waveguide~\cite{Waveguide,PAr_Ding}. 
$\kappa \triangleq \tfrac{2\pi}{\lambda_f}$ is the wavenumber, and $\left\Vert \bm{\eta}_{n,l}\left(x_{n,l}\right)-\bm{\eta}_{n}^{\mathrm{W}}\right\Vert= x_{n,l}$ 
is the distance from waveguide $n$ to PA $a_{n,l}$.
Since the transmitting power is equally radiated by each PA along the waveguide, the amplitude of the transmitted signal is multiplied by $1/\sqrt{L}$. 

We consider high-frequency bands such as millimeter-wave (mmWave) or terahertz (THz) in this work. 
Owing to the severe path-loss and shadowing at the high-frequency bands, 
the power gains of non-line-of-sight (NLoS) paths are typically negligible~\cite{Channel_LoS}.  
Hence, the channel vector can be approximated by the LoS-dominant model. 
Let $\mathbf{h}_{k,n}^{H}\left(\mathbf{x}_{n}\right) \in\mathbb{C}^{1\times L}$ denote the 
channel vector between PAs $\mathcal{L}_{n}$ and user $k$, which is determined 
by the pinching deployment $\mathbf{x}_{n}$. 
The $l$-th element of channel vector $\mathbf{h}_{k,n}^{H}\left(\mathbf{x}_{n}\right)$ 
indicates the channel from PA $a_{n,l}$ to user $k$ at location $\bm{\eta}_{k}^{\mathrm{U}}$,  
which is given by the geometric free-space spherical wavefront model~\cite{Spherical_Channel}:
\begin{equation}\label{channel_ori}
        h_{k,n,l}^{H}\left(x_{n,l}\right) = 
        \frac{\sqrt{\beta}e^{-i\kappa r\left(x_{n,l}, \bm{\eta}_{k}^{\mathrm{U}}\right) }}
        {r\left(x_{n,l}, \bm{\eta}_{k}^{\mathrm{U}}\right)},
\end{equation}
where $\kappa=2\pi/\lambda_{f}$ is the wave-domain number, and $\lambda_{f}$ is the wavelength. 
The constatnt $\beta ={c}/{\left(4\pi f_{c}\right)}$ depends on the speed of light $c$ and the carrier frequency $f_{c}$, 
which represents the reference channel gain at a range of unit distance (i.e., $1$ meter). 
Moreover, $\sqrt{\beta}/{r\left(x_{n,l}, \bm{\eta}_{k}^{\mathrm{U}}\right)}$ 
denotes the channel gain coefficient, 
and $r\left(x_{n,l}, \bm{\eta}_{k}^{\mathrm{U}}\right) = \left\Vert \bm{\eta}_{k}^{\mathrm{U}}-\bm{\eta}_{n,l}\left(x_{n,l}\right)\right\Vert$ 
is the distance between PA $a_{n,l}$ and user $k$, which is computed by 
\begin{equation}\label{distance}
    r\left(x_{n,l},\!\bm{\eta}_{k}^{\mathrm{U}}\right)=
    \sqrt{\left(x_{n,l}-x_{k}^{\mathrm{U}}\right)^{2}
    +\left(y_{n,l}^{\mathrm{PA}}-y_{k}^{\mathrm{U}}\right)^{2}+ \left(h^{\mathrm{PA}}\right)^{2}}.
\end{equation}

Let $\mathbf{h}_{k}^{H}\left(\mathbf{X}\right) \!=\! 
\big[\mathbf{h}_{k,1}^{H}\!\left(\mathbf{x}_{1}\right), \mathbf{h}_{k,2}^{H}\!\left(\mathbf{x}_{2}\right), 
\dots, \mathbf{h}_{k,N}^{H}\!\left(\mathbf{x}_{N}\right)\big] \in \mathbb{C}^{1\times M}$ stack 
the channel vectors from all the PAs to user $k$. 
Hence, the received signal at user $k$ through the channels manipulated by PAs can be compactly expressed by
\begin{equation}
    y_{k}\!=\!\underset{\text{desired signal}}{\underbrace{\overset{\text{pinching beamforming}}
    {\overbrace{\mathbf{h}_{k}^{H}\!\!\left(\mathbf{X}\right)\mathbf{G}\left(\mathbf{X}\right)}}
    \!\mathbf{d}_{k}\widetilde{s}_{k}}}\!+\!\underset{\text{multi-user interference}}
    {\underbrace{\sum_{k'\ne k}\!\!\overset{\text{pinching beamforming}}
    {\overbrace{\mathbf{h}_{k}^{H}\!\!\left(\mathbf{X}\right)\!\mathbf{G}\!\left(\mathbf{X}\right)}}
    \!\!\mathbf{d}_{k'}\widetilde{s}_{k'}}}
    \!+\!n_{k}.
\end{equation} 
The effective gains of user $k$ can be given by 
\begin{equation}\label{effective_gain}
    E_{k} \! = \!\frac{\beta}{L}\left|\sum\limits_{n\in\mathcal{N}}
    \sum\limits_{l\in\mathcal{L}_{n}}\frac{
    e^{-i\kappa \left(r\left(x_{n,l},\bm{\eta}_{k}^{\mathrm{U}}\right)+n_{\mathrm{eff}}x_{n,l}\right)}}
    {r\left(x_{n,l},\bm{\eta}_{k}^{\mathrm{U}}\right)}d_{n,k}
    \right|^{2}.
\end{equation} 
Furthermore, the interference experienced by user $k$ caused by the transmitted signal of user $k'$ can be represented by 
\begin{equation}\label{interference_term}
    I_{k, k'} \!= \!\frac{\beta}{L}\left|\sum\limits_{n\in\mathcal{N}}
    \sum\limits_{l\in\mathcal{L}_{n}}\frac{e^{-i\kappa \left(r\left(x_{n,l}, \bm{\eta}_{k}^{\mathrm{U}}\right)
    +n_{\mathrm{eff}}x_{n,l}\right)}}
    {r\left(x_{n,l}, \bm{\eta}_{k}^{\mathrm{U}}\right)}
    d_{n,k'}\!\right|^{2}. 
\end{equation}
Therefore, the signal-to-interference-and-noise ratio (SINR) of user $m$ is given by
\begin{equation}
    \mathrm{SINR}_{k}\!=\! 
    \frac{\left|\mathbf{h}_{k}^{H}\!\left(\mathbf{X}\right)\!\mathbf{G}\!\left(\mathbf{X}\right)\!\mathbf{d}_{k}\right|^{2}}
    {\sum\limits_{k'\ne k}\!\left|\mathbf{h}_{k}^{H}\!\left(\mathbf{X}\right)\!\mathbf{G}\!\left(\mathbf{X}\right)\!\mathbf{d}_{k'}\right|^{2}
    \!\!+\!\sigma^{2}}
    \!\!=\!\! \frac{E_{k}}{\sum\limits_{k'\ne k} I_{k, k'} \!+\! \sigma^{2}}.
\end{equation} 
To enhance SINR of users, the transmit beamforming and the pinching beamforming should be 
jointly designed to achieve constructive interference  of radio waves for improving effective gains $E_{k}$ in \eqref{effective_gain}, 
while achieving destructive interference for mitigating multi-user interference $I_{k,k'}$ in \eqref{interference_term}.

\subsection{Problem Formulation and Conversion}
The key objective of this paper is to maximize the sum rate of users in PASS  
by jointly optimizing the transmit beamforming and the pinching beamforming. 
This is formulated as the following sum rate maximization problem: 
\begin{subequations}\label{P0}
    \begin{align*}
        \text{(P0)} ~ &  \max_{\mathbf{X}, \mathbf{D}}
        ~ \sum_{k\in\mathcal{K}}\log_{2} \left(1\!+\! \frac{\left|\mathbf{h}_{k}^{H}\!\left(\mathbf{X}\right)\!\mathbf{G}\!\left(\mathbf{X}\right)\!\mathbf{d}_{k}\right|^{2}}
        {\sum\limits_{k'\ne k}\left|\mathbf{h}_{k}^{H}\left(\mathbf{X}\right)\mathbf{G}\left(\mathbf{X}\right)\mathbf{d}_{k'}\right|^{2}+\sigma^{2}}\right) \tag{\ref{P0}{a}} \\
        \text{s.t.} ~
        & \text{(C1): } x_{n,l} \!-\! x_{n,l-1} \geqslant \Delta_{\min},  ~ \forall 1 < l \leqslant L, \forall n\in\mathcal{N}, \tag{\ref{P0}{b}} \label{P0_x} \\   
        & \text{(C2): }0 \leqslant x_{n,l} \leqslant S_{\mathrm{x}}, ~ \forall l \in \mathcal{M}, \tag{\ref{P0}{c}} \label{P0_waveguide} \\        
        & \text{(C3): }\sum_{k\in\mathcal{K}} \left \Vert \mathbf{d}_{k} \right \Vert_{2}^{2}\leqslant P, \tag{\ref{P0}{d}} \label{P0_power}
    \end{align*}
\end{subequations}
where (C1) ensures the minimum antenna space $\Delta_{\min}$ to avoid mutual coupling, 
(C2) guarantees the location of each PA does not exceed the maximum length of the connected waveguide, 
and (C3) denotes the maximum transmitting power of the BS. 
It is worth noting that (P0) is a highly non-convex optimization problem with strongly coupled variables. 
Conventionally, we can solve (P0) by decomposing it into decoupled problems and invoking convex approximation techniques to find suboptimal solutions. 
In the remaining sections, we will first come up with an MM-PDD algorithm based on conventional optimization theory to achieve stationary solution of (P0). 
Then, by revisiting the KKT solution structure of the beamforming problem, a novel learning algorithm will be developed 
to further deal with the coupled transmit and pinching beamforming optimization.

\section{Optimization Based Joint Transmit and Pinching Beamforming}

To solve problem (P0), we first reformulate the SINR expression into a more tractable form 
based on WMMSE. 
Then, an MM-PDD optimization algorithm is developed, which recasts the coupling constraints 
into the augmented Lagrangian function, thus alternatively updating multi-block variables in a BCD style. 
During the block variable update, the nonconvex component is relaxed by majorization minimization, 
where the Lipschitz gradient surrogate is employed as the block surrogate for stable convergence. 

\subsection{WMMSE-based Reformulation}
Assume $\widetilde{s}_{k}$ and $n_{k}$ are independent.  
Therefore, the mean square error (MSE) of user $k$ can be given by ~\cite{WMMSE}
\begin{multline}\label{MSE}
    \hspace{-0.5em}\!e_{k}\!\left(\mathbf{X},\mathbf{D},\mathbf{v}\right)\!
    \triangleq\!\mathbb{E}\left[\left|\widehat{y}_{k}\!-\!\widetilde{s}_{k}\right|^{2}\right] 
    \!\!=\!\! \mathbb{E}\left[\left(v_{k}y_{k}-\widetilde{s}_{k}\right)^{H}\!\!\left(v_{k}y_{k}\!-\!\widetilde{s}_{k}\right)\right] \!\!=\!\!
    \\
    \!\!\sum_{i\in\mathcal{K}}\!\left|v_{k} \mathbf{h}_{k}^{H}\!\!\left(\mathbf{X}\right)\!\mathbf{G}\!\!\left(\mathbf{X}\right)
    \!\mathbf{d}_{i}\right|^{2}
    \!+\sigma^{2}\!\left|v_{k}\right|^{2}\!+\!1
    \!-\!2\mathrm{Re}\!\left\{v_{k}\mathbf{h}_{k}^{H}\!\!\left(\mathbf{X}\right)\!\mathbf{G}\!\left(\mathbf{X}\right)\!
    \mathbf{d}_{k}\!\right\}\!,
\end{multline}
where $\widehat{y}_{k} = v_{k} y_{k}$ is the estimated signal received by user $k$ and $v_{k}$ denotes the channel equalizer. 
\begin{lemma}\label{lemma:WMMSE}
The optimal solution of the constrained sum rate maximization problem (P1) is equivalent to the following WMMSE problem:
\begin{subequations}\label{P1}
    \begin{align*}
        \text{(P1)} ~  \min_{\mathbf{X},\mathbf{D}, \mathbf{v}, \bm{\alpha}}&
        \sum_{k=1}^{K}\left(\alpha_{k}e_{k}\left(\mathbf{X},\mathbf{D},v_{k}\right)-\log_{2}\alpha_{k}\right), 
        \tag{\ref{P1}{a}} \label{P1_obj}\\
        \text{s.t.} ~
        & \text{(C1) - (C3)}, \tag{\ref{P1}b}
    \end{align*}
\end{subequations}
where $\alpha_{k}$ denotes the weighting factor. 
By fixing other variables, the optimal solution of $\alpha_{k}$ and $v_{k}$ can be given by 
\begin{equation}\label{MMSE_detection}
    v_{k}^{opt}\!=\! J_{k}^{-1} \mathbf{d}_{k}^{H} \mathbf{G}_{k}^{H}\left(\mathbf{X}\right)\mathbf{h}_{k}\left(\mathbf{X}\right), 
    \forall k \in \mathcal{K},
\end{equation}
\begin{equation}\label{weighting_opt}
\alpha_{k}^{opt}\!\!=\!\!\left(1-\!
    \mathbf{d}_{k}^{H}\!\mathbf{G}^{H}\!\!\left(\mathbf{X}\right)\mathbf{h}_{k}\!\left(\mathbf{X}\right)J_{k}^{-1}
    \mathbf{h}^{H}\!\left(\mathbf{X}\right)\!\mathbf{G}\!\left(\mathbf{X}\right)\!\mathbf{d}_{k}\right)^{-1}\!,
\end{equation}
where $J_{k}\triangleq\sum\limits_{i\in\mathcal{K}}\!\mathbf{h}_{k}^{H}\!\left(\mathbf{X}\right)\!\mathbf{G}\!\left(\mathbf{X}\right)
\mathbf{d}_{i}\mathbf{d}_{i}^{H}\!\mathbf{G}^{H}\!\left(\mathbf{X}\right)\!\mathbf{h}_{k}\!\left(\mathbf{X}\right)\!+\!\sigma^{2}$ 
denotes the covariance of signal $y_{k}$ received by user $k$.
\begin{proof}
See Appendix \ref{proof:WMMSE}.
\end{proof}
\end{lemma}

The above WMMSE-based reformulation converts the fractional SINR into a more tractable form. 
We could follow the BCD method to optimize (P1) if it is multi-convex with respect to (w.r.t.) each block variable $\mathbf{D}$ and $\mathbf{x}$. 
However, the complex exponential components and the reciprocals of distances in \eqref{MSE} render the non-convexity of (P1) w.r.t. $\mathbf{x}$, 
which is strongly coupled with $\mathbf{D}$. 
To solve this nonconvex coupled problem, we invoke the PDD theory with a majorization minimization procedure, thus ensuring the stable convergence to a critical point.

\begin{figure*}[b]
    \centering
    \setcounter{equation}{\numexpr\getrefnumber{L_NC}\relax}
    \begin{subequations}\label{AL_antgain_relax_x}
        \begin{align*}
            L_{k,n,l}^{\mathrm{AL}}\left(x_{n,l}\right)
            &=\!\underset{\text{$\triangleq L_{k,n,l}^{\mathrm{DC}}\left(x_{n,l}\right)$, a D.C. function if $\Omega_{k,n,l}<0$}}
            {\underbrace{\overset{\text{$\triangleq L_{k,n,l}^{\mathrm{conv}}\left(x_{n,l}\right)$, convex over $x_{k,n,l}$}}
            {\overbrace{\frac{1}{2\rho}\!\left(\left|u_{k,n,l}^{2}\right|\!+\!\kappa^{2}\right)\!\left(x_{n,l}\!-\!x_{k}^{\mathrm{U}}\right)^{2}
            \!+\!\frac{1}{2\rho}\kappa^{2}n_{\mathrm{eff}}^{2}x_{n,l}^{2}\!-\!\kappa\!\left(\lambda_{k,n,l}^{\mathrm{\theta}}
            \!+\!\frac{\theta_{k,n,l}}{\rho}\right)\!n_{\mathrm{eff}}x_{n,l}}}\!+\!L_{k,n,l}^{\mathrm{CC}}\!\left(x_{n,l}\right)}}
            \!+\!L_{k,n,l}^{\mathrm{NC}}\left(x_{n,l}\right), \label{AL_antgain_expand_x}\tag{\ref{AL_antgain_relax_x}{a}}
            \\
            \!&\leqslant\! L_{k,n,l}^{\mathrm{MM}}\left(x_{n,l}\right)L_{k,n,l}^{\mathrm{conv}}\!\left(x_{n,l}\right) \!\triangleq\! 
            \!+\! \widehat{L}_{k,n,l}^{\!\mathrm{CC}}\!\left(x_{n,l}\right) \!+\! \widehat{L}_{k,n,l}^{\mathrm{NC}}\!\left(x_{n,l}\right),
            \label{AL_antgain_MM_x}\tag{\ref{AL_antgain_relax_x}{b}}
            \\
            \widehat{L}_{k,n,l}^{\!\mathrm{CC}}\!\left(x_{n,l}\right) \!& =\! 
            L_{k,n,l}^{\!\mathrm{CC}}\!\left(x_{n,l}^{(t-1)}\right)\!
            \!+\!\nabla_{\!x}L_{k,n,l}^{\mathrm{CC}}\!\!\left(x_{n,l}^{(t-1)}\right)\!\!\left(x_{n,l}\!-\!x_{n,l}^{(t-1)}\!\right)\!
            \!=\! L_{k,n,l}^{\!\mathrm{CC}}\!\left(x_{n,l}^{(t-1)}\right)\!
            \!+\!\tfrac{\Omega_{k,n,l}\left(x_{n,l}^{(t-1)}\!-\!x_{k}^{\mathrm{U}}\right)}
            {\left(x_{n,l}^{(t-1)}-x_{k}^{\mathrm{U}}\right)^{2}\!+\!\psi_{k,n,l}^{2}}\!\!\left(x_{n,l}\!-\!x_{n,l}^{(t-1)}\!\right)\!, 
            \label{L_CC_x_MM}\tag{\ref{AL_antgain_relax_x}{c}}
            \\
            \widehat{L}_{k,n,l}^{\mathrm{NC}}\!\left(x_{n,l}\right) \! & \triangleq \! 
            \tfrac{x_{n,l}^{3}\!+\!\left(x_{k}^{\mathrm{U}}\right)^{2}\!+\!
            \big(2\psi_{k,n,l}^{2}
            \!+\!\left(x_{n,l}^{(t-1)}\!-\!x_{k}^{\mathrm{U}}\right)^{2}\!-\!4x_{k}^{\mathrm{U}}x_{n,l}^{(t-1)}\big)x_{n,l}\!+\!
            2x_{k}^{\mathrm{U}}\left(x_{n,l}^{(t-1)}\right)^{2}}{2\sqrt{\left(x_{n,l}^{(t-1)}-x_{k}^{\mathrm{U}}\right)^{2}+\psi_{k,n,l}^{2}}}
            \!\geqslant\!L_{k,n,l}^{\mathrm{NC}}\!\left(x_{n,l}\right)
            \!\triangleq \!x_{n,l}\sqrt{\left(x_{n,l}\!-\!x_{k}^{\mathrm{U}}\right)^{2}\!+\!\psi_{k,n,l}^{2}}.
            \label{L_NC_x_MM}\tag{\ref{AL_antgain_relax_x}{d}}
        \end{align*}
    \end{subequations}
\end{figure*}

\subsection{The Proposed MM-PDD Algorithm}
We first transform (P1) into a more tractable augmented Lagrangian (AL) problem, 
and then establish Lipschitz gradient surrogate for convex relaxation based on the MM method. 
To this end, we newly introduce auxiliary variable 
$\theta_{k,n,l}=\kappa\left(r\left(x_{n,l},\bm{\eta}_{k}^{\mathrm{U}}\right) + n_{\mathrm{eff}}x_{n,l}\right)$ to indicate the PASS-modified phase of user $k$'s signal, 
and $u_{k,n,l} = \phi e^{-i\theta_{k,n,l}}/r\left(x_{n,l},\bm{\eta}_{k}^{\mathrm{U}}\right)$ to signify the equivalent pinching beamforming coefficient of user $k$'s signal 
radiated by PA $a_{n,l}$, 
where  $\phi\triangleq \sqrt{{\beta}}/\sqrt{L}$ is a constant. 
This leads to the following equality constraints:
\setcounter{equation}{\numexpr\getrefnumber{weighting_opt}\relax}
\begin{subequations}\label{eq_antgain}
    \begin{equation}
        b_{k,n,l}^{u}\!\triangleq\!u_{k,n,l}\sqrt{\left(x_{n,l}\!-\!x_{k}^{\mathrm{U}}\right)^{2}\!+\!\psi_{k,n,l}^{2}}\!
        -\! \phi e^{-i\theta_{k,n,l}}\!=\!0, 
    \end{equation}
    \begin{equation}
        b_{k,n,l}^{\theta}\!\triangleq\!\theta_{k,n,l}\!-\!\kappa\!\left(\!\sqrt{\left(x_{n,l}\!-\!x_{k}^{\mathrm{U}}\!\right)^{2}
        \!+\!\psi_{k,n,l}^{2}}\!+\!n_{\mathrm{eff}}x_{n,l}\!\right)\!=\!0,
\end{equation} 
\end{subequations}
where $\psi_{k,n,l}\!\triangleq\! \sqrt{\left(y_{n,l}^{\mathrm{PA}}\!-\!y_{k}^{u}\right)^{2}\!+\!\left(h^{\mathrm{PA}}\right)^{2}}$ is fixed for each channel realization. 
Moreover, we define auxiliary variable 
$q_{k, k'}= \mathbf{h}_{k}^{H}\left(\mathbf{X}\right)\mathbf{G}\left(\mathbf{X}\right) \mathbf{d}_{k'}$ as 
the path response of user $k$ to receive the signal of user $k'$.  
From the above definitions, $q_{k,k'}$ satisfies the equality constraint: 
\begin{equation}\label{eq_effgain}
    b_{k,k'}^{q} \triangleq q_{k,k'} - \mathbf{u}_{k}^{\top} \bm{\Sigma} \mathbf{d}_{k'} = 0, ~ \forall{k, k'} \in \mathcal{K},    
\end{equation}
where $\mathbf{u}_{k}^{\top} = \left[\mathbf{u}_{k,1}^{\top}, \mathbf{u}_{k,2}^{\top}, \dots, \mathbf{u}_{k,N}^{\top} \right]\in\mathbb{C}^{1\times M}$ 
stacks the pinching beamforming coefficients for all the PAs, 
and $\mathbf{u}_{k,n}^{\top}=\left[u_{k,n,1}, u_{k,n,2}, \dots, u_{k,n,L}\right]\in\mathbb{C}^{1\times L}$ 
is the pinching beamforming vector of user $k$ using PAs in subset $\mathcal{L}_{n}$. 
Moreover, $\bm{\Sigma} = \mathrm{blkdiag} \left(\mathbf{1}_{L\times 1}, \mathbf{1}_{L\times 1}, \dots, \mathbf{1}_{L\times 1}\right) \in \mathbb{R}^{M\times N}$ 
is a block diagonal matrix. 

\begin{figure*}[b]
    \centering
    \setcounter{equation}{21}
    \begin{subequations}\label{L_NC_Thetb_Relax}
        \begin{equation}\label{Lipschitz_surrogate_Theta}
            L_{k,n,l}^{\mathrm{ex}}\left(\theta_{k,n,l}\right)
            \! \leqslant \! \widehat{L}_{k,n,l}^{\mathrm{ex}}\left(\theta_{k,n,l}\right)\!
            \!=\!L_{k,n,l}^{\mathrm{ex}}\left(\theta_{k,n,l}^{(t-1)}\right)+\nabla_{\theta}L_{k,n,l}^{\mathrm{ex}}\left(\theta_{k,n,l}^{(t-1)}\right)\left(\theta_{k,n,l}-\theta_{k,n,l}^{(t-1)}\right)
            \!+\!\left| \tfrac{\varrho_{k,n,l}^{\theta}}{2}\left(\theta_{k,n,l}-\theta_{k,n,l}^{(t-1)}\right)\right| ^{2},
        \end{equation}
        \begin{equation}\label{Lipschitz_surrogate_derivative_Theta}
            \nabla_{\theta}L_{k,n,l}^{\mathrm{ex}}\left(\theta\right)
            \!=\!\phi\mathrm{Re}\left\{ \lambda_{k,n,l}^{u}+\tfrac{u_{k,n,l}r_{k,n,l}}{\rho}\right\} \sin\theta+
            \phi\mathrm{Im}\left\{ \lambda_{k,n,l}^{u}+\tfrac{u_{k,n,l}r_{k,n,l}}{\rho}\right\} \cos\theta.
        \end{equation}
    \end{subequations}
\end{figure*}

By penalizing and dualizing equalities into the objective function, the AL function of (P1) can be constructed by:
\setcounter{equation}{\numexpr\getrefnumber{eq_effgain}\relax}
\begin{align*}
        \text{(P2)} ~ \min_{\mathbf{x},\mathbf{D}, \bm{\alpha}, \mathbf{v},  \mathbf{S}, \mathbf{U}, \bm{\theta}, \mathbf{Q}}~
        &  \sum_{k=1}^{K}\!\left(\alpha_{k}e_{k}\!\left( \mathbf{Q},v_{k}^{opt}\right)\!\!-\!\log_{2}\!\alpha_{k}\right)  \!
        \notag \\ & \hspace{-6em}\! 
        \!+\!\frac{1}{2\rho} \!\sum_{k\in\mathcal{K}}\!\!\Bigg(\!\!\left\Vert \mathbf{B}_{k}^{u} \!\!
        +\!\! \rho\bm{\lambda}_{k}^{u}\right\Vert^{2} 
        \!+\!\left\Vert \mathbf{B}_{k}^{\theta} \!\!+ \!\!\rho\bm{\lambda}_{k}^{\theta}\right\Vert^{2} 
        \!+\!\left\Vert \mathbf{b}_{k}^{q} \!\!+\!\! \rho\bm{\lambda}_{k}^{q}\right\Vert^{2}\!\!\Bigg),
        \notag\\
        &  \hspace{-4em}\text{s.t.} ~
         \text{(C1) - (C3)}, \notag
\end{align*}
where $\mathbf{Q} = \left\{q_{k,k'}\right\} \in\mathbb{C}^{K\times K}$ 
and $\mathbf{U} =\left[\mathbf{u}_{1},\mathbf{u}_{2}, \dots, \mathbf{u}_{K}\right]\in\mathbb{R}^{M \times K}$.  
$\mathbf{B}_{k}^{u}=\left\{b_{k,n,l}^{u}\right\}$ and $\mathbf{B}_{k}^{\theta}=\left\{b_{k,n,l}^{\theta}\right\}$, 
stack the residuals of the equality constraints in \eqref{eq_antgain},
and $\mathbf{b}_{k}^{q}=\left\{b_{k',k}^{q}\right\}$ stacks the residuals of equality constraint \eqref{eq_effgain}. 
Moreover, $\bm{\lambda}_{k}^{u}\in\mathbb{C}^{N\times L}$, $\bm{\lambda}_{k}^{\theta}\in\mathbb{C}^{N\times L}$, 
and $\bm{\lambda}_{k}^{q}\in\mathbb{C}^{K\times 1}$ are dual variables 
corresponding to \eqref{eq_antgain} and \eqref{eq_effgain}, respectively.
By fixing $v_{k}^{opt}$, the MSE is rewritten as a convex function of w.r.t. $q_{k,k'}$:
\begin{equation}
    e_{k}\!\!\left(\mathbf{Q},v_{k}^{opt}\right)
    \!\!=\!\!\left|v_{k}^{opt}\right|^{2}\!\!\left(\sum_{k'\in\mathcal{K}}\!\left|q_{k,k'}\right|^{2}\!
    +\!\sigma^{2}\!\right)\!
    \!+1-\!2\mathrm{Re}\!\left\{ v_{k}^{opt}q_{k,k}\right\}.
\end{equation}
Now, the nonconvexity of problem (P2) only lies in the AL term 
$\left\Vert \mathbf{B}_{k}^{u} + \rho\bm{\lambda}_{k}^{u} \right\Vert^{2} + 
\left\Vert \mathbf{B}_{k}^{\theta} + \rho\bm{\lambda}_{k}^{\theta} \right\Vert^{2}$ 
w.r.t. $\mathbf{x}$ and $\bm{\theta}$, which is the sum of terms 
\begin{equation}\label{L_NC}
    \begin{split}
        &L_{k,n,l} \!=\!
        \!\left|u_{k,n,l}\sqrt{\!\left(x_{n,l}\!-\!x_{k}^{\mathrm{U}}\right)^{2}\!\!+\!\psi_{k,n,l}^{2}}
        \!-\!\!\phi e^{-i\theta_{k,n,l}}\!\!+\!\rho\lambda_{k,n,l}^{u}\right|^{2}
        \!+\!\\&
        \!\left(\!\!\theta_{k,n,l}\!-\!\kappa\!\sqrt{\!\left(x_{n,l}\!-\!x_{k}^{\mathrm{U}}\right)^{2}
        \!+\!\psi_{k,n,l}^{2}}\!-\!\kappa n_{\mathrm{eff}}x_{n,l}\!+\!\rho\lambda_{k,n,l}^{\theta}\!\!\right)^{\!\!2}\!.
    \end{split}
\end{equation}
We construct convex surrogate functions over the block variables $\mathbf{x}$ and $\bm{\theta}$ during the block update of PDD, respectively. 

Regarding $x_{n,l}$, the AL function \eqref{L_NC} can be rewritten as function $L_{k,n,l}^{\mathrm{AL}}\left(x_{n,l}\right)$  
 defined in \eqref{AL_antgain_expand_x} in the bottom of the next page, 
which consists of a difference of convex (D.C.) function $L_{k,n,l}^{\mathrm{DC}}\big(x_{n,l} \big)$, 
and a nonconvex component 
$L_{k,n,l}^{\mathrm{NC}}\big(x_{n,l}\big)\!\triangleq\!\frac{\kappa^{2}n_{\mathrm{eff}}}{\rho}x_{n,l}\sqrt{\big(x_{n,l}\!-\!x_{k}^{\mathrm{U}}\big)^{2}+\psi_{k,n,l}^{2}}$. 
We construct the MM surrogate of $L_{k,n,l}^{\mathrm{AL}}\left(x_{n,l}\right)$ in \eqref{AL_antgain_expand_x} as follows. 
\textit{First}, for the D.C. function $L_{k,n,l}^{\mathrm{DC}}\big(x_{n,l} \big)$, we can observe that $L_{k,n,l}^{\mathrm{conv}}\big(x_{n,l} \big)$ is convex over $x_{n,l}$, 
and $L^{\mathrm{CC}}\!\left(x_{n,l}\right) \!\triangleq \!
\Omega_{k,n,l}\sqrt{\left(x_{n,l}\!-\!x_{k}^{\mathrm{U}}\right)^{2}\!+\!\psi_{k,n,l}^{2}}$ is a concave component 
when the coefficient $\Omega_{k,n,l}\triangleq\mathrm{Re}\big\{ u_{k,n,l}^{H}
\big(\lambda_{k,n,l}^{u}\!-\!\tfrac{\phi e^{-i\theta_{k,n,l}}}{\rho}\big)\big\} 
\!+\!\big(\lambda_{k,n,l}^{s}\!-\!\tfrac{r_{k,n,l}}{\rho}\big)$ is negative. 
We handle the D.C. component using the concave-convex procedure (CCCP), 
which is a special case of MM for D.C. programming~\cite{CCCP_NIPS, MM_TSP}. 
CCCP utilizes the first-order Taylor expansion $\widehat{L}^{\mathrm{CC}}\!\left(x_{n,l}\right)$  
as the tight upper bound of the concave component $L^{\mathrm{CC}}\!\left(x_{n,l}\right)$ if $\Omega_{k,n,l}\leqslant 0$, 
as presented in \eqref{L_CC_x_MM} in the bottom of this page.
\textit{Secondly}, for the nonconvex component $L_{k,n,l}^{\mathrm{NC}}\big(x_{n,l}\big)$, 
from Jensen's inequality we have $\sqrt{x}\leqslant\sqrt{x^{(t-1)}}+\frac{x-x^{(t-1)}}{2\sqrt{x^{(t-1)}}}\leqslant\frac{x+x^{(t-1)}}{2\sqrt{x^{(t-1)}}}$. 
Applying the above inequality we have 
\setcounter{equation}{\numexpr\getrefnumber{AL_antgain_relax_x}\relax}
\begin{equation}\label{MM_x_sqrt_upbound}
\sqrt{\!\left(x_{n,l}\!-\!x_{k}^{\mathrm{U}}\right)^{2}\!+\!\psi_{k,n,l}^{2}}\!\leqslant
\!\tfrac{\left(x_{n,l}\!-\!x_{k}^{\mathrm{U}}\right)^{2}\!
\!+\!\left(x_{n,l}^{\!(t-1)}\!-\!x_{k}^{\mathrm{U}}\right)^{2}\!
\!+\!2\psi_{k,n,l}^{2}}{2\sqrt{\!\left(x_{n,l}^{(t-1)}\!-\!x_{k}^{\mathrm{U}}\right)^{2}\!+\!\psi_{k,n,l}^{2}}}\!.
\end{equation}
Substituting \eqref{MM_x_sqrt_upbound} into $L_{k,n,l}^{\mathrm{NC}}\big(x_{n,l}\big)$ and exploiting the 
first-order Taylor expansion 
$-x_{n,l}^{2}\!\leqslant\!
-\!\big(x_{n,l}^{(t-1)}\big)^{2}\!-\!2x_{n,l}^{(t-1)}\big(x_{n,l}-x_{n,l}^{(t-1)}\big)
\!=\!\big(x_{n,l}^{(t-1)}\big)^{2}-2x_{n,l}^{(t-1)}x_{n,l}$, 
the MM surrogate of $L_{k,n,l}^{\mathrm{NC}}\big(x_{n,l}\big)$ can be derived, as shown in \eqref{L_NC_x_MM}.

Regarding $\theta_{k,n,l}$, given a fixed PA-user distance $r_{k,n,l} = r(x_{n,l},\bm{\eta}_{k}^{\mathrm{U}})$, the objective function 
$L_{k,n,l}^{\mathrm{NC}}$ involves the complex exponential component $L^{\mathrm{ex}}\left(\theta_{k,n,l}\right) \!=\! -\phi\mathrm{Re}\big\{ \big(\lambda_{k,n,l}^{u}
\!+\!\frac{u_{k,n,l}r_{k,n,l}}{\rho}\big)e^{i\theta_{k,n,l}}\big\}$,  
which periodically switch between convex and concave functions over the feasible range of $\theta_{k,n,l}\in\mathbb{R}$. 
To convert this intractable nonconvex component, we construct Lipschitz gradient surrogate based on the MM method~\cite{Lipschitz_surrogate}, 
which can be defined as follows.
\begin{definition}[Lipschitz gradient surrogate]
    The Lipschitz gradient surrogate of a function $f\left(\bm{\theta}\right)$ is defined as
    \begin{multline}\label{eq:Lipschitz_surrogate}
        \mathcal{S}\left(\bm{\theta},\bm{\theta}^{(t-1)}\right)
            =f\left(\bm{\theta}^{(t-1)}\right) +
            \\ \!\!\left(\!\nabla_{\bm{\theta}}f\!\left(\bm{\theta}^{(t-1)}\!\right)\!\right)^{\!T}\!\!\!\!\left(\bm{\theta}\!-\!\bm{\theta}^{(t-1)}\!\right)
            \!\!+\!\!\frac{\varrho}{2}\left\Vert \bm{\theta}\!-\!\bm{\theta}^{(t-1)}\!\right\Vert^{2}\!,
    \end{multline}
    where $\varrho$ is the Lipschitz gradient constant that ensures $\left\Vert \nabla_{\bm{\theta}} f\left(\bm{\theta}_{1}\right) 
    -\nabla_{\bm{\theta}}f\left(\bm{\theta}_{2}\right) \right\Vert\leqslant \varrho \Vert \bm{\theta}_{1}-\bm{\theta}_{2}\Vert$, $\forall \bm{\theta}_{1}, \bm{\theta}_{2}$. 
\end{definition}

The Lipschitz gradient surrogate $\mathcal{S}\left(\bm{\theta},\bm{\theta}^{(t-1)}\right)$ 
establishes an MM upper bound of $f\left(\bm{\theta}\right)$ when the 
gradient $\nabla_{\bm{\theta}}f\left(\bm{\theta}^{(t-1)}\!\right)$ is $\varrho$-Lipschitz continuous~\cite{Lipschitz_surrogate,MM_TSP}. 
Compared to the first-order Taylor expansion, Lipschitz gradient surrogate exploits an extra second-order regularization term, 
which achieves a strict convexity approximation and thus improves the convergence guarantees.  
To apply MM, we introduce the following lemma to construct Lipschitz gradient surrogate of 
the exponential term $L^{\mathrm{ex}}\left(\theta_{k,n,l}\right)
=-\phi\mathrm{Re}\left\{ \left(\lambda_{k,n,l}^{u}+\frac{u_{k,n,l}r_{k,n,l}}{\rho}\right)e^{i\theta_{k,n,l}}\right\}$. 
\begin{lemma}\label{lemma:Lipschitz}
   Given $a, \theta \in\mathbb{R}$ and $c\in\mathbb{C}$, function $-\mathrm{Re}\left\{ ce^{i\left(a\theta\right)}\right\}$ has Lipschitz-continuous gradient w.r.t. $\theta$,  
    and the Lipschitz constant is given by $\varrho^{\theta} = a^2 |c|.$
    \begin{proof} 
        See Appendix \ref{proof:Lipschitz}.
    \end{proof}
\end{lemma}

From the above Lemma, the Lipschitz gradient surrogate of 
$L_{k,n,l}^{\mathrm{ex}}\left(\theta_{k,n,l}\right)$ 
can be constructed by \eqref{L_NC_Thetb_Relax}, which is defined in the bottom of this page.
Furthermore, the corresponding Lipschitz constant can be given by 
\setcounter{equation}{22}
\begin{equation}
    \varrho_{k,n,l}^{\theta}=\phi\left|\lambda_{k,n,l}^{u}+\frac{u_{k,n,l}r_{k,n,l}}{\rho}\right|. 
\end{equation}

Based on the PDD method~\cite{PDD}, we can now solve (P1) by alternatively optimizing the following four subproblems in a BCD manner in the inner loop, 
while updating the dual variables and penalty factor in the outer loop.

\noindent(i) \textit{Subproblem w.r.t. $\left\{\mathbf{D},\mathbf{Q}\right\}$}: 
The transmit beamforming $\mathbf{D}$ and the path response $\mathbf{Q}$ are jointly 
optimized by solving the following constrained optimization problem:
\begin{subequations}\label{P_D}
    \begin{align}
        \min_{\mathbf{Q},\mathbf{D}}~ & 
        \sum_{k=1}^{K}\alpha_{k}e_{k}\left(\mathbf{Q},v_{k}^{opt}\right)
        \!+\!\frac{1}{2\rho}\left\Vert \mathbf{Q}
        \!\!-\!\!\mathbf{U}^{\top}\bm{\Sigma}\mathbf{D}\!+\!\rho\bm{\lambda}^{q}\right\Vert ^{2}\!\!,
        \\&
        \text{s.t.} \quad \text{(C3)}.
    \end{align}
\end{subequations}
The optimal solution of $q_{k,k'}$ can be obtained by 
\begin{equation}
    q_{k,k'}^{*}=\frac{2v_{k}\delta_{k,k'}-\lambda_{k,k'}^{q}+\frac{1}{\rho}\mathbf{u}_{k}^{\top}\bm{\Sigma}\mathbf{d}_{k}}{2|v_{k}|^{2}+\frac{1}{\rho}}, 
    ~\forall k, k' \in \mathcal{K},
\end{equation}
where $\delta_{k,k'}=1$ if $k=k'$ and $\delta_{k,k'}=0$ otherwise. 

\noindent(ii) \textit{Subproblem w.r.t. $\mathbf{X}$}: Combining the convex components and the block surrogate function, 
the pinching deployment $\mathbf{X}$ can be updated by solving 
\begin{align}\label{P_x}
        \min_{\mathbf{X}} ~
        & \frac{1}{2\rho} \sum_{k\in\mathcal{K}}\left\Vert \bm{\theta}_{k} 
        \!-\! \kappa \left(\mathbf{s}_{k} \!+\! n_{\mathrm{eff}} \mathbf{X}\right) \right\Vert^2 
        \!+\!  \sum_{k\in\mathcal{K}}\!\sum_{n\in\mathcal{N}}\!\sum_{l\in\mathcal{L}_{n}}
        \!\widehat{L}_{k,n,l}^{\mathrm{DC}}\left(x_{n,l}\right)\!
         \notag
        \\&
        \text{s.t.} \quad \text{(C1), (C2)}. 
\end{align}
The optimal solution of this convex quadratic programming problem can be solved by the interior point method~\cite{Interior_Point}.

\noindent(iii) \textit{Subproblem w.r.t. $\mathbf{U}$}: 
The effective pinching beamforming coefficients $\mathbf{U}$ can be obtained by solving the unconstrained convex quadratic optimization problem: 
\begin{equation}\label{P_U}
    \min_{\mathbf{U}}
    \!\sum_{k\in\mathcal{K}}\!\left(\left\Vert \mathbf{R}_{k}\mathbf{u}_{k}
    \!+\!\bm{\zeta}_{k}\right\Vert ^{2}\!+\!\left\Vert \mathbf{q}_{k}\!+\!\rho\bm{\lambda}_{k}^{q}
    \!-\!\mathbf{u}_{k}^{\top}\bm{\Sigma}\mathbf{D}\right\Vert ^{2}\!\right)\!,
\end{equation}
where $\mathbf{R}_{k}\!\!=\!\!\text{diag}\big(r_{k,1,1},r_{k,1,2},\dots,r_{k,N,L}\big)
\!\in\!\mathbb{R}^{M\times M}$, and $r_{k,n,l}=r\left(x_{n,l},\bm{\eta}_{k}^{\mathrm{U}}\right)$. 
Moreover, 
$\zeta_{k,n,l}=\rho\lambda_{k,n,l}^{u}-\phi e^{-i\theta_{k,n,l}}$, and vector 
$\bm{\zeta}_{k}=\left[\zeta_{k,1,1},\zeta_{k,1,2},\dots,\zeta_{k,N,L}\right]^{\top}\in\mathbb{C}^{M\times1}$. 
% $\widetilde{\bm{\lambda}}_{k}^{u}=\big[\lambda_{k,1,1},\lambda_{k,1,2},\dots,\lambda_{k,N,L}\big]\in\mathbb{C}^{M\times1}$. 
% and $\widetilde{\bm{\theta}}_{k}^{u}=\big[\theta_{k,1,1},\theta_{k,1,2},\dots,\theta_{k,N,L}\big]\in\mathbb{C}^{M\times1}$.
Hence, the closed-form solution of $\mathbf{u}_{k}$, $\forall k \in \mathcal{K}$, can be derived as
\begin{equation}\label{Sol_MMPPD_U}
    \mathbf{u}_{k}^{*}\!=\!\left(\mathbf{R}_{k}^{H}\mathbf{R}_{k}\!+\!
    \bm{\Sigma}\mathbf{D}\mathbf{D}^{H}\!\bm{\Sigma}^{H}\!\right)^{-1}\!\!\left(\bm{\Sigma}
    \mathbf{D}\!\left(\mathbf{q}_{k}\!+\!\rho\bm{\lambda}_{k}^{q}\right)^{H}\!\!\!-\!\mathbf{R}^{H}\!\bm{\zeta}_{k}\!\right)\!.
\end{equation}
% Using $\frac{\partial L_{\lambda}(\mathbf{U})}{\partial\mathbf{U}}=\mathbf{0}_{M\times K}$, the closed form solution of 
% $\mathbf{U}$ can be given by

\noindent(iv) \textit{Subproblem w.r.t. $\bm{\theta}$}: 
The signal phases $\bm{\theta}$ and the PA-user distance $\mathbf{S}$ can be updated by solving the unconstrained convex optimization problem: 
\begin{multline}\label{P_Theta}
    \min_{\bm{\theta},\mathbf{S}} ~
    \sum_{n,l}\sum_{k}\Bigg[\frac{1}{2\rho}\!\left(\theta_{k,n,l}\!-\!\kappa\!\left(r_{k,n,l}\!+\!n_{\mathrm{eff}}x_{n,l}\right)
    +\rho\lambda_{k,n,l}^{\theta}\right)^{2} 
    \\ + \!\!\nabla_{\theta}L_{k,n,l}^{\mathrm{ex}}\!\left(\theta_{k,n,l}^{(t-1)}\right)\!\!
    \theta_{k,n,l}\!+\!\frac{\varrho_{k,n,l}^{\theta}}{2}
    \!\left(\theta_{k,n,l}\!-\!\theta_{k,n,l}^{(t-1)}\right)^{2}
    \!\Bigg]\!,
\end{multline}
where the optimal solution can be given by 
\begin{equation}\label{Sol_theta}
    \theta_{k,n,l}^{\star}\!\!=\!\!\frac{\varrho_{k,n,l}^{\theta}\theta_{k,n,l}^{(t-1)}\!\!-\!\!
    \lambda_{k,n,l}^{\theta}\!\!+\!\!
    \frac{\kappa}{\rho}\!\left(r_{k,n,l}\!+\!n_{\mathrm{eff}}x_{n,l}\!\right)\!\!
    -\!\!\nabla_{\!\!\theta}L^{\!\mathrm{ex}}
    \!\!\left(\!\theta_{k,n,l}^{(t-1)}\!\right)\!}{\varrho_{k,n,l}^{\theta}\!+\!\frac{1}{\rho}}\!.
\end{equation}

\begin{algorithm}[!t]
    \caption{MM-PDD algorithm Using Lipschitz gradient surrogate}
    \label{alg:MMPDD}
    \begin{algorithmic}[1]
    \REQUIRE Users' locations $\bm{\eta}^{\mathrm{U}}$ and feed points' locations $\bm{\eta}^{\mathrm{W}}$.
    % \ENSURE Global optimal solution $\bm{\rho}^{*}$
    % -------------------------------------------------
    % \item[] {\textit{//*~\textbf{Initialization}}:}
    \STATE Initialize primal variables $\mathbf{x}$, $\mathbf{D}$, $\mathbf{Q}$, and $\mathbf{U}$.  
    \STATE Initialize dual variables $\bm{\lambda}^{(0)} = \mathbf{0}$ and tolerance threshold.
    \FOR{$i = 1, 2, \dots, I_{\max}$}
        \item[] {//*~\textit{\textbf{MM-BCD procedure}}:}
        \WHILE{convergence criterion is not satisfied}
        \STATE Update $v_{k}$ and $\alpha_{k}$ by \eqref{MMSE_detection} and \eqref{weighting_opt}, $\forall k \in \mathcal{K}$. 
        \STATE Update $\left\{\mathbf{D},\mathbf{Q}\right\}$ by solving \eqref{P_D} while fixing remaining variables. 
        \STATE Update $\mathbf{x}$ by \eqref{P_x} while fixing remaining variables.
        \STATE Update $\mathbf{U}$ by \eqref{Sol_MMPPD_U}, and then update $\bm{\theta}$ by \eqref{Sol_theta} via Lipschitz gradient surrogate \eqref{Lipschitz_surrogate_Theta}.% while fixing remaining variables. 
        \ENDWHILE
        \IF{$\left\Vert\mathbf{B}^{(i)}\right\Vert_{\infty} \leqslant \varepsilon^{(i)} \!\triangleq \! 
        \gamma \left\Vert\mathbf{B}^{(i-1)}\right\Vert_{\infty}$}
            \STATE Update $\bm{\lambda}^{(i+1)}\leftarrow \bm{\lambda}^{(i)} + \frac{1}{\rho^{(i)}} \mathbf{B}^{(i)}$, $\rho^{(i+1)} \leftarrow \rho^{(i)}$.
        \ELSE
            \STATE Update $\rho^{(i+1)} \leftarrow \varsigma \rho^{(i)}$, $\bm{\lambda}^{(i+1)}\leftarrow \bm{\lambda}^{(i)}$.
        \ENDIF
        \STATE Algorithm terminates if $\left\Vert\mathbf{B}^{(i)} \right\Vert_{\infty} \leqslant \varepsilon$. 
    \ENDFOR
    
    \ENSURE $\mathbf{X}, \mathbf{D}$.
    
    \end{algorithmic}
\end{algorithm}

\subsection{Algorithm Design and Theoretical Analysis}
Algorithm \ref{alg:MMPDD} summarizes the iterative procedure of MM-PDD\footnote{In each PDD iteration $i$, the residual tolerance $\varepsilon^{(i)}=\gamma \|\mathbf{B}^{(i)}\|_\infty $ 
controls whether to update the dual variables or to decrease the penalty parameter (by $\rho^{(i+1)}\leftarrow \varsigma \rho^{(i)}$), 
where $0 < \gamma, \varsigma < 1$ enforces equality residuals $\|\mathbf{B}^{i}\|_{\infty}\to 0$ as $i$ grows. 
We empirically set $\gamma=0.9$ and $\varsigma=0.85$, which achieves an acceptable convergence speed without sacrificing the solution quality. 
}.

The computational complexity of the proposed MM-PDD algorithm can be analyzed as follows. In each inner iteration, 
the updates of $\{v_k,\alpha_k\}$ require evaluating the effective channels $\mathbf h_k(\mathbf X)\mathbf G(\mathbf X)$, $\forall k$, 
which scales linearly with the number of pinching antennas as $\mathcal{O}(KM)$. 
The update of $\{\mathbf D,\mathbf Q\}$ involves solving a convex quadratic subproblem in $\mathbf D$, whose time complexity using interior-point methods 
is given by $\mathcal{O}(N^{3.5}K^{3.5})$, and updating $\mathbf{Q}$ incurs a time complexity $\mathcal{O}(KM+K^2)$. 
The update of the pinching deployment $\mathbf{X}$ relies on solving a convex quadratic program with $NL$ variables, 
which typically dominates the iteration cost with $\mathcal{O}(N^{3.5}L^{3.5})$. 
The update of $\mathbf{U}$ requires $K$ independent $M \times M$ linear system inversions, with overall computational complexity $O(K M^3)$. 
Finally, the updates of $\bm{\theta}$ is closed-form and only costs $\mathcal{O}(KM)$. 
Therefore, the time complexity of MM-PDD with $N_{\mathrm{it}}$ inner iterations can be given by 
$\mathcal{O}\big(N_{\mathrm{it}}\big((K^{3.5}+L^{3.5}) N^{3.5} + M^{3}K + K^2 + 2KM\big)\big)$.

The developed MM-PDD is mathematically guaranteed to converge to a stationary solution, as proven below.
\begin{lemma}\label{lemma:MMPDDconvergence}
    % Suppose the objective and constraints are locally Lipschitz continuous. 
    \textbf{Algorithm \ref{alg:MMPDD}} converges to a stationary solution (i.e., a KKT point) of problem (P0). 
    \begin{proof}
        See Appendix \ref{proof:lemma:MMPDDconvergence}.
    \end{proof}
\end{lemma}

Despite the strict theoretical guarantees, obtaining a high quality solution $\left\{\mathbf{X}^{*}, \mathbf{D}^{*}\right\}$ by MM-PDD is highly complex, 
and is prone to getting stuck in local optima. 
This motivates us to further develop the learning-based approach.

\section{Learning-Based Joint Transmit and Pinching Beamforming}
In this section, we introduce preliminaries on learning-based methods and present the fundamentals of the proposed KDL. 
Then, a KDL-Transformer is further developed for joint transmit and pinching beamforming.
\subsection{Preliminaries on Learning to Optimize}
The non-convex coupled sum rate optimization can be written in the following form: 
\begin{subequations}\label{P_general}
    \begin{align}
        \left(\mathcal{P}_{\mathbf{z}}\right) \min_{\mathbf{X}, \mathbf{D}} ~
        & F_{0}\left(\mathbf{X}, \mathbf{D};\mathbf{z}\right), \tag{\ref{P_general}a} \label{P_general_obj} \\
        \text{s.t.} ~
        & f_{i}\left(\mathbf{X}\right) \!\leqslant\! 0, ~ \forall i \!\in\! \mathcal{F}_{X}, %\tag{\ref{P_general}b} \label{P_general_ineq_x} \\
        ~ f_{j}\left(\mathbf{D}\right) \!\leqslant\! 0, ~ \forall j \!\in\! \mathcal{F}_{D}. \tag{\ref{P_general}b} \label{P_general_ineq_D}
    \end{align}
\end{subequations}
where  $\mathbf{z}$ is the system parameter vector (e.g., CSI), and 
$\mathbf{X}$ and $\mathbf{D}$ denote the optimization variable blocks, which are coupled in 
the objective function $F_{0}\left(\mathbf{X}, \mathbf{D};\mathbf{z}\right)$. 
Moreover, $\mathcal{F}_{X}$ and $\mathcal{F}_{D}$ denote the sets of $F_{x}$ and $F_{D}$ inequality constraints, respectively.
% The optimization objective is defined as $F_{0}\left(\mathbf{X};\mathbf{z}\right)\triangleq 
%         \phi\left(\mathbf{X};\mathbf{z}\right) + \sum_{n=1}^{N} \eta\left({\mathbf{X}_{n}};\mathbf{z}\right)$, 
% where $\phi\left(\mathbf{X};\mathbf{z}\right)$ is a non-separable function, and 
% $\eta\left(\mathbf{X}_{n}; \mathbf{z}\right)$ is a separable function over block variables $\mathbf{X}_{n}$, $\forall n$.  
All the functions in \eqref{P_general} are differentiable and smooth, but may be nonlinear and non-convex. 

Generally, the optimization-based algorithm alternatively optimizes the coupled variables in $\mathcal{P}_{\mathbf{z}}$ through an iterative process and relies on convex surrogate. 
To overcome their shortcomings, the concept of learning to optimize (L2O) has emerged~\cite{L2O}, which trains ML model parameters $\mathbf{w}$ 
to learn the mapping from input parameters $\mathbf{z}$ to desired solutions $\widehat{\mathbf{X}}$,  i.e., 
$\widehat{\mathbf{X}} \triangleq \widehat{\mathbf{X}}_{\mathbf{w}}\left(\mathbf{z}\right)$. 
Thus, optimization variables can be simultaneously predicted without requiring sophisticated alternating computations or iterative procedures. 
Despite the attractive promises of L2O, how to efficiently deal with nonconvex coupled optimization 
and to achieve comparative performance with conventional optimization method are still open challenges. 
For this purpose, existing L2O methods can be roughly classified into two categories, i.e., 
end-to-end learning and unfolding-based learning. 
\subsubsection{End-to-end learning}
L2O can train the learning model parameters $\mathbf{w}$ in an end-to-end fashion, 
which directly takes the observed system parameters $\mathbf{z}$ as the input of ML model,  
and then predicts solutions $\mathbf{X}$ and $\mathbf{D}$. 
As ML model (e.g., deep neural networks) can approximate arbitrary nonconvex functions, 
the end-to-end learning can mimic high-quality solution structure based on supervised learning.
The model parameters $\mathbf{w}$ are trained to minimize the mean-square-error (MSE) loss function:
\begin{equation}
    \min_{\mathbf{w}} \left\Vert \mathbf{X}^{*} - \widehat{\mathbf{X}}_{\mathbf{w}}\left(\mathbf{z}\right)\right\Vert^{2}
    + \left\Vert \mathbf{D}^{*} - \widehat{\mathbf{D}}_{\mathbf{w}}\left(\mathbf{z}\right)\right\Vert^{2},
\end{equation}
where $\widehat{\mathbf{X}}_{\mathbf{w}}\left(\mathbf{z}\right)$ and 
$\widehat{\mathbf{D}}_{\mathbf{w}}\left(\mathbf{z}\right)$ denote the solution predicted by ML model parameters $\mathbf{w}$ for the given input parameters $\mathbf{z}$. 
However, supervised learning requires a large-scale dataset of pre-solved instances (samples), 
which is impractical for a nonconvex coupled optimization problem, especially when the system complexity (e.g., number of users/antennas) increases.
Therefore, the unsupervised learning has become a viable option, which trains the 
model parameters $\mathbf{w}$ by employing the optimization objective as the loss function:
\begin{equation}
    % \begin{split}
        \mathbf{w}=\!\mathop{\arg\min}_{\mathbf{w}} 
        F_{0}\left(\widehat{\mathbf{X}}_{\mathbf{w}}\left(\mathbf{z}\right), \widehat{\mathbf{D}}_{\mathbf{w}}\left(\mathbf{z}\right)\right).
    %     \\&
    %     =\!-\mathbb{E}_{\mathbf{z}\in\mathcal{H}}
    %     \left[\sum_{k\in\mathcal{K}}\log_{2}
    %     \left(1+\gamma_{k}\left(\widehat{\mathbf{X}}_{\mathbf{w}}\left(\mathbf{z}\right), 
    %     \widehat{\mathbf{D}}_{\mathbf{w}}\left(\mathbf{z}\right)\right)\right)\right]\!\!.
    % \end{split}
\end{equation}
However, as unsupervised learning relies on gradient descent, it is also prone to saddle points and local minimums 
when tackle coupled and non-convex optimization problem. 

\subsubsection{Deep unfolding-based optimization}
To guide the ML model training through the optimization theory and to seek theoretical guarantees,  
deep unfolding further proposes to unrolls an iterative optimization algorithm by a sequence of neural network (NN) layers~\cite{DeepUnfolding,DeepUnfoldingWMMSE}, 
e.g., unfolding WMMSE for transmit beamforming optimization.  
The NN layers are trained to mimic the iterative computation procedure of conventional algorithms.
However, performing such unrolling procedure requires closed-form solutions to be fully understood for coupled optimization problem. 
Furthermore, the design principle of deep unfolding inherits the iterative update nature of conventional optimization algorithm, 
which may still limit its computational efficiency in practical applications.

\subsection{The Proposed KKT-guided Dual Learning Approach}
To overcome the aforementioned challenges, we propose a novel theory-guided end-to-end learning architecture 
in this paper, namely KDL, which can efficiently tackle the coupled beamforming optimization problem.  
Unlike previous end-to-end learning or unfolding-based learning, KDL directly trains ML models to approximate the KKT solution structure of some block optimization variables, 
whose explicit KKT-conditioned solution structure is easy to be derived by Lagrangian duality method. 
KDL learns dual variables that are required to reconstruct KKT solutions of these block optimization variables. 
The dual variables can be jointly inferred with the remaining optimization variables that lack closed-form KKT solution structure  
using the end-to-end unsupervised learning. 
Therefore, KDL can capitalize on the white-box solution structure obtained from the optimization theory, 
while inheriting the efficient execution natrue of the data-driven method for high-quality solution exploration. 
% The key idea of KDL is to decompose the coupled optimization problem over multiple block variables.
% For block variables $\mathbf{D}$ that has an explicit KKT solution structure, % but desirable solutions w.r.t. block variables $\mathbf{X}$ may be difficult to obtain.  
% KDL allows the machine learning model to jointly predict the dual variables $\bm{\lambda}$ and the remaining primal variables $\mathbf{X}$. 

\subsubsection{Fundamentals of KDL}
To illustrate the fundamental principles of KDL, we consider that block variables $\mathbf{D}$ possess explicit KKT solutions, 
while the  closed-form solutions of remaining block variables $\mathbf{X}$ are difficult to obtain. 
By fixing block variables $\mathbf{X}$, the Lagrangian function w.r.t. block variable $\mathbf{D}$ is given by
\begin{equation}
    L_{\mathrm{Dual}}\left(\mathbf{D},\bm{\lambda};\mathbf{X},\mathbf{z}\right)\!=\!F_{0}\left(\mathbf{D};\mathbf{X},\mathbf{z}\right)
    +\sum_{j\in\mathcal{F}_{\mathrm{D}}}\lambda_{j}f_{j}\left(\mathbf{D},\bm{\lambda}\right),
\end{equation}
where $\bm{\lambda}\in\mathbb{R}^{F_{D}\times 1}$ 
denotes the dual variables for all inequality constraints in \eqref{P_general_ineq_D}.  
The KKT conditions corresponding to problem ($\mathcal{P}_{\mathbf{z}}$) can be expressed as~\cite{ConvexOpt}
\begin{equation}\label{P_general_KKT}
    \!\!\left\{\!\!\!
        \begin{array}{l}
            \text{(i) }  f_{j} \left(\mathbf{D};\mathbf{z}, \mathbf{X}\right) \geqslant 0, 
            ~ \forall j \in \mathcal{F}_{D}.
            \\
            \text{(ii) } \bm{\lambda} \geqslant \mathbf{0}, 
            \\\text{(iii) }
            \!\nabla_{\mathbf{D}}F_{0}\left(\mathbf{D};\mathbf{X},\mathbf{z}\right)
            \!+\!\sum\limits _{j\in\mathcal{F}_{D}}\lambda_{j}\nabla_{\mathbf{D}}f_{j}\left(\mathbf{D}\right)
            \!=\!\mathbf{0},
            \\
            \text{(iv) }
            \lambda_{j}f_{j}\left(\mathbf{D},\bm{\lambda}\right)=0, ~ \forall j \in \mathcal{F}_{D}.
        \end{array}
    \right.
\end{equation}
where (i)-(ii) denote the primal and dual feasibility conditions, (iii) is the stationary condition (i.e., 
the first-order optimality $\nabla_{\mathbf{X}} L_{\mathrm{AL}}=\mathbf{0}$), 
and (iv) denotes the complementary slackness conditions. 
% The main difference from standard KKT conditions lies in the complementary slackness conditions (iv). 
% The complementary slackness conditions are given by  
% $\lambda_{j}f_{j}\left(\mathbf{D},\bm{\lambda}\right)=0$, $\forall j \in\mathcal{F}_{D}$.  
After rearranging the equality constraints in \eqref{P_general_KKT}, one can obtain the KKT solutions of block variables $\mathbf{D}$ by solving the following system: 
\begin{equation}\label{KKT_condition_system}
    \bm{\varphi}\!\left(\mathbf{D},\bm{\tau},\bm{\lambda},\bm{\mu}\right)\!\!\triangleq\!\!\left[\!\begin{array}{c}
        \nabla_{\mathbf{D}}F_{0}\!\left(\mathbf{D};\mathbf{X},\mathbf{z}\right)\!
        +\!\mathbf{J}_{\mathbf{f},\mathbf{D}}^{\top}\bm{\lambda}\\
        \mathrm{diag}\left(\bm{\lambda}\right)\mathbf{f}\left(\mathbf{D}\right)
        \end{array}\!\right]\!=\!\left[\!\begin{array}{c}
        \mathbf{0}\\
        \mathbf{0}
        \end{array}\!\right],
\end{equation}
% where $\mathbf{J}_{\mathbf{X}}=\left[\frac{\partial z_{1}\left(\mathbf{X},\bm{\tau};\mathbf{z}\right)}{\partial\mathbf{X}},
% \frac{\partial z_{2}\left(\mathbf{X},\bm{\tau};\mathbf{z}\right)}{\partial\mathbf{X}},\dots,
% \frac{\partial z_{\widetilde{F}_{\mathrm{E}}}\left(\mathbf{X},\bm{\tau};\mathbf{z}\right)}{\partial\mathbf{X}}\right]\in 
% \mathbb{R}^{N_{\mathrm{x}}\times\widetilde{F}_{\mathrm{E}}}$ 
% and $\mathbf{J}_{\bm{\tau}}=\left[\frac{\partial z_{1}\left(\mathbf{X},\bm{\tau};\mathbf{z}\right)}{\partial\bm{\tau}},
% \frac{\partial z_{2}\left(\mathbf{X},\bm{\tau};\mathbf{z}\right)}
% {\partial\bm{\tau}},\dots,\frac{\partial z_{\widetilde{F}_{\mathrm{E}}}\left(\mathbf{X},\bm{\tau};\mathbf{z}\right)}
% {\partial\bm{\tau}}\right]\in\mathbb{R}^{J_{\mathrm{IN}}\times\widetilde{F}_{\mathrm{E}}}$
% are Jacobian matrices of constraint functions $\mathbf{z}\left(\mathbf{X},\bm{\tau};\mathbf{z}\right)$ 
% over $\mathbf{X}$ and $\bm{\tau}$, respectively.
where $\mathbf{J}_{\mathbf{f},\mathbf{D}}^{\top}
=\left[\nabla_{\mathbf{D}}f_{1}\left(\mathbf{D}\right),\nabla_{\mathbf{D}}f_{2}\left(\mathbf{D}\right),\dots,
\nabla_{\mathbf{D}}f_{F_{D}}\left(\mathbf{D}\right)\right]$ 
is the Jacobian matrix of constraint functions $\mathbf{f}\left(\mathbf{D};\mathbf{z}\right)$ 
over $\mathbf{D}$.

Conventionally, the closed-form KKT solution of $\mathbf{D}$ can be derived 
when the equality system \eqref{KKT_condition_system} has a relatively simple structure. 
The resulting KKT solution of $\mathbf{D}$ would be expressed as a function of the input system parameters $\mathbf{z}$ and the remaining decision variables $\mathbf{X}$, i.e., 
$\mathbf{D}^{\mathrm{KKT}} = F^{\mathrm{KKT}}\left(\bm{\lambda}, \mathbf{X}, \mathbf{z}\right)$.
The key idea of KDL is to \textbf{predict dual variables $\bm{\lambda}$ by learning and directly reconstruct KKT solutions using the data-driven method.}
Instead of predicting all primal variables by block-box method or unfolding the computationally-expensive iterative optimization procedure, KDL 
forces ML to straightforwardly explore the KKT solution structure of 
$\mathbf{D}^{\mathrm{KKT}}$ using
\begin{equation}
    \widehat{\mathbf{D}}_{\mathbf{w}}^{\mathrm{KKT}} 
    = F^{\mathrm{KKT}}\left(\widehat{\bm{\lambda}}_{\mathbf{w}}\left(\mathbf{z}\right), 
    \widehat{\mathbf{X}}_{\mathbf{w}}\left(\mathbf{z}\right)\right).
\end{equation}
Therefore, KDL not only enables the guidance of white-box structure from optimization theory,  
but also retains the computation efficiency of data-driven and end-to-end learning.

\begin{remark}
We may not obtain the closed-form solution of all variable blocks in most coupled optimization problem. 
When conventional optimization algorithm is adopted, the remaining variable block $\mathbf{X}$ 
that does not have closed-form solution needs to be computed by solving \eqref{KKT_condition_system} using iterative numerical method, 
such as interior point method and Newton method~\cite{Interior_Point}. 
However, this typically requires intensive computations when the number of users/antennas increases.
Hence, to maintain a high computation efficiency during execution, 
KDL will jointly learn the dual variables $\bm{\lambda}$ 
and the remaining block variables $\mathbf{X}$ using the date-driven ML model. 
\end{remark}

\subsubsection{KDL Based Joint Transmit Beamforming and Pinching Beamforming}
To apply the proposed KDL paradigm for joint transmit and pinching beamforming optimization, we first provide specific analysis in this part.
The KKT solution structure of the original sum rate optimization problem is hard to obtain even using alternative optimization. 
However, the WMMSE based reformulated problem (P1) provides insights into the KKT solution structure of $\mathbf{D}$ when the remaining variables are fixed, 
as analyzed as follows. 
Specifically, the Lagrangian dual function of (P1) can be given by
\begin{equation}
    \begin{split}
        L_{\mathrm{dual}}(\mathbf{D},\lambda)
        \!=\!\sum_{k}\alpha_{k}e_{k}\left(\mathbf{X},\mathbf{D}\right)\!+\!\lambda\left(\sum_{k}\|\mathbf{d}_{k}\|^{2}\!-\!P\right).
    \end{split}
\end{equation}
For ease of notations, the equivalent channels manipulated by pinching beamforming is defined as
\begin{equation}
    \widetilde{\mathbf{h}}_{k}^{H}\left(\mathbf{X}\right) = \mathbf{h}_{k}^{H}\left(\mathbf{X}\right) \mathbf{G} \left(x\right),
~ \forall k \in\mathcal{K}.
\end{equation}
Using the definition of MSE, i.e.,  
$\sum\limits_{k\in\mathcal{K}} \!\alpha_k e_k \left(\mathbf{X},\mathbf{D}\right) \!=\!
\sum\limits_{k\in\mathcal{K}} \!\alpha_k \!\bigg( \!\sum\limits_{i\in\mathcal{K}} \!
|v_k \widetilde{\mathbf{h}}_{k}^{H}\!\!\left(\mathbf{X}\right) \!\mathbf{d}_i|^2 \!+\! 
\sigma^2 |v_k|^2 \!+\! 1
\!-\! 2\mathrm{Re} \left\{ v_k  \widetilde{\mathbf{h}}_{k}^{H}\!\left(\mathbf{X}\right) \mathbf{d}_k \right\}
\bigg)$, 
the first-order optimality condition of $L_{\mathrm{dual}}\left(\mathbf{D},\lambda\right)$ w.r.t. $\mathbf{d}_{k}$ 
can be expressed as
\begin{equation}
    \frac{\partial L_{\mathrm{dual}}}{\partial \mathbf{d}_k} \!=\!
    \sum_{i\in\mathcal{K}}\!\alpha_{i}|v_{i}|^{2}\widetilde{\mathbf{h}}_{i}\!\left(\mathbf{X}\right)\!
    \widetilde{\mathbf{h}}_{i}^{H}\!\left(\mathbf{X}\right)\!\mathbf{d}_{k}
    \!-\!\alpha_{k}v_{k}^{H}\widetilde{\mathbf{h}}_{k}\!\left(\mathbf{X}\right)\!+\!\lambda\mathbf{d}_{k}
    \!=\! \mathbf{0},
    % \frac{\partial L_{\mathrm{dual}}}{\partial\mathbf{d}_{k}}\!=\!\sum_{i\in\mathcal{K}}\!\alpha_{i}|v_{i}|^{2}
    % \mathbf{G}^{H}\!\left(\mathbf{X}\right)\!\mathbf{h}_{i}\!\left(\mathbf{X}\right)\!\mathbf{h}_{i}^{H}\!\left(\mathbf{X}\right)\!
    % \mathbf{G}\left(\mathbf{X}\right)\!\mathbf{d}_{k}\!
    % \\-\!\alpha_{k}v_{k}^{H}\mathbf{G}^{H}\!\left(\mathbf{X}\right)\mathbf{h}_{k}\!\left(\mathbf{X}\right)+\!\lambda\mathbf{d}_{k}\!=\!\mathbf{0},
\end{equation}
\begin{equation}\label{KKT_D_ori}
    \mathbf{d}_{k}^{\mathrm{KKT}}\!\!\!=\!\!\left(\mathbf{I}_{N}
    \!\!+\!\!\sum_{i}\!\frac{\alpha_{i}|v_{i}|^{2}}{\lambda}
    \!\widetilde{\mathbf{h}}_{i}\!\left(\mathbf{X}\right)\!\widetilde{\mathbf{h}}_{i}^{H}\!\left(\mathbf{X}\right)\!\right)^{-1}\!\!\!\!\frac{\alpha_{k}v_{k}^{H}}{\lambda}
    \!\widetilde{\mathbf{h}}_{k}\!\left(\mathbf{X}\right)\!.
\end{equation}
Considering the phase rotation equivalency of transmit beamforming $\mathbf{d}_{k}$, 
when we rotate the phase of $\mathbf{d}_{k}$ into $e^{j \theta} \mathbf{d}_{k}$, 
the achievable data rate will not be changed since $\left|\widetilde{\mathbf{h}}_{i}\left(\mathbf{X}\right)\mathbf{d}_{k}\right|^{2} = \left|e^{j\theta}\widetilde{\mathbf{h}}_{i}\left(\mathbf{X}\right)\mathbf{d}_{k}\right|^{2}$, $\forall i \in \mathcal{K}$. 
Using this property,  we can always rotate the phase of $\mathbf{d}_{k}^{\mathrm{KKT}}$ in \eqref{KKT_D_ori} to make $v_{k}^{H}$ a real and positive value 
(i.e., $v_{k}^{H}= v_{k} = \left|v_{k}\right|$) without harming the system performance. 
This means that the KKT solution structure of $\mathbf{d}_{k}$ can be simplified into 
\begin{equation}\label{KKT_D}
    \mathbf{d}_{k}^{\mathrm{KKT}}=\mu_{k}\left(\mathbf{I}_{N}+\sum_{i}\lambda_{i}\widetilde{\mathbf{h}}_{i}\left(\mathbf{X}\right)
    \widetilde{\mathbf{h}}_{i}^{H}\left(\mathbf{X}\right)\right)^{-1}\widetilde{\mathbf{h}}_{k}\left(\mathbf{X}\right),
\end{equation}
where $\lambda_{k}=\alpha_{k}|v_{k}|^{2}/\lambda > 0$ is the weighted dual variable of each user $k$, and $\mu_{k}=\lambda_{k}/v_{k}$ is a certain (real) scalar that 
controls the power allocation. 
\begin{remark}
    As observed above, the KKT solution of WMMSE-based transmit beamforming optimization shares a highly similar structure 
     with the optimal beamforming structure in~\cite{BF_Emil}, which is derived 
by the KKT solution of a reformulated power minimization problem. This confirms its effectivity from the theoretical aspect. 
\end{remark}
Based on KDL,  we only need to predict dual variables $\bm{\lambda} = \left[\lambda_{1}, \lambda_{2}, \dots, \lambda_{K}\right]$, 
 power allocation coefficients $\bm{\mu} = \left[\mu_{1}, \mu_{2}, \dots, \mu_{K}\right]$, and remaining block variables $\mathbf{X}$ using the ML model $\mathcal{W}\left(\mathbf{z}\right)$:
\begin{equation}
    \left[{\bm{\lambda}}, {\bm{\mu}}, {\mathbf{X}}\right] = \mathcal{W}\left(\mathbf{z}\right),
\end{equation}
where $\mathbf{z} = \left[\mathbf{X}^{\mathrm{U}}, \mathbf{y}^{\mathrm{U}}\right] \in \mathbb{R}^{2K\times 1}$ denotes the vectorized CSI characterized by users' locations, 
and $\mathcal{W}\left(\cdot\right)$ denotes the function of ML model (e.g., deep neural networks). 
Then, the KKT solution in \eqref{KKT_D} can be reconstructed by
\begin{equation}\label{DBF_KKT_matrix}
    \mathbf{D}^{\mathrm{KKT}}\!\!=\!\!\text{diag}\!\left({\bm{\mu}}\right)\!\!
    \left(\mathbf{I}_{N}\!+\!\widetilde{\mathbf{H}}\!\left({\mathbf{X}}\right)
    \!\text{diag}\!\left({\bm{\lambda}}\right)\!\widetilde{\mathbf{H}}^{H}\!\!\left({\mathbf{X}}\right)\!\right)^{\!\!-1}
    \!\!\widetilde{\mathbf{H}}\!\left({\mathbf{X}}\right)\!,
\end{equation}
where $\widetilde{\mathbf{H}}\left(\mathbf{X}\right) \triangleq \mathbf{H}^{H}\left(\mathbf{X}\right)\mathbf{G}\left(\mathbf{X}\right)$ is the effective channel matrix.
In this way, KDL can benefit from both the optimization-guided structure and the iteration-free data-driven learning. 
\begin{remark}
    The dimension of output parameters can be reduced by KDL from $N \times K \times 2$ complex transmit beamforming coefficients to only $2 K$ dual variables, 
    which enables fast convergence of the learning algorithm.
\end{remark}

\subsection{The Proposed KDL-Transformer Algorithm}
We implement the proposed KDL based on Transformer, which give rises to a KDL-Transformer algorithm in this part. 
Specifically, the CSI-to-beamforming optimization problem is regarded as a sequence-to-sequence learning task. 
The CSI of users is considered as the input sequence, while transmit beamforming and pinching beamforming coefficients are determined using the output sequence. 
Transformer is known as one of the most core techniques of LLM to model  intricate structural data~\cite{Trnasformer}, such as nature language, visual, and even multi-modal data. 
Compared to conventional deep learning techniques, the developed Transformer can capture the long-distance dependence between elements in the input and output sequences. 
To be more specific, it can dynamically aggregate the context information of PASS by modelling: 
\textbf{(i) inter-PA and inter-user dependencies} that models the spatial distribution and interactions of pinches/users for constructive interference and destructive interference, and 
\textbf{(ii) CSI-beamforming  dependencies} that maps various user distributions into efficient transmit and pinching beamforming solutions.   
By customizing the self-attention and cross-attention mechanisms to learn these relationships, 
the developed KDL-Transformer algorithm can achieve efficient effective gain enhancement and interference suppression in PASS.

\begin{figure}[!t]
    \centering
    \includegraphics[width=0.38\textwidth]{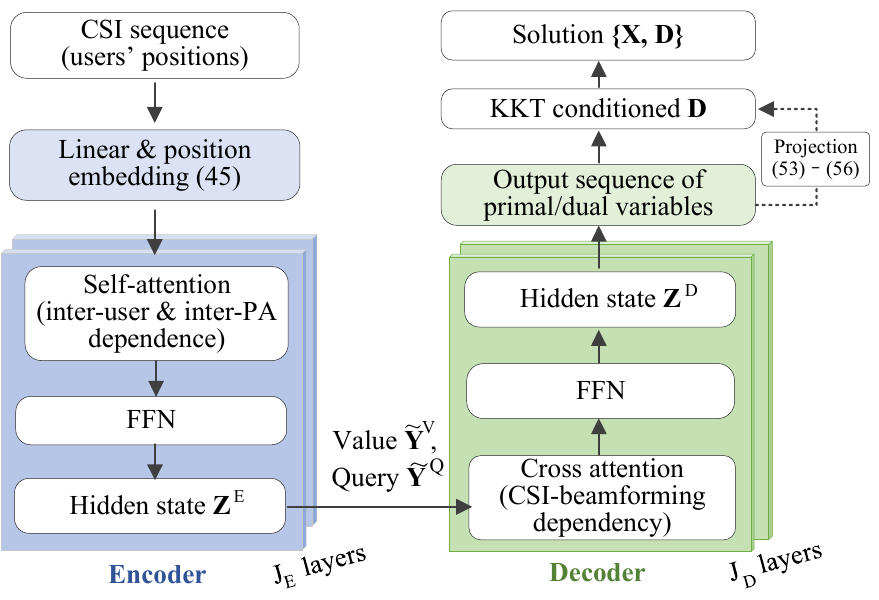}
    \caption{Diagram of the proposed KDL-Transformer.}
    \label{fig_KDL}
\end{figure}

% We develop the following customized Transformer structure to realize above goals. 
Following existing LLM structure, the developed Transformer consists of a $J_{\mathrm{E}}$-layer encoder $\mathcal{E}$ and a $J_{\mathrm{D}}$-layer decoder $\mathcal{D}$, 
as shown in Fig. \ref{fig_KDL}.
The Transformer structure is tailored to utilize bidirectional attention mechanisms, thus modelling global dependence in both the encoder and the decoder:
(i) Both the encoder and decoder adopt bidirectional self-attention mechanisms, which characterizes 
the inter-user dependence in the input sequence and the inter-PA dependence in the output sequence, respectively. 
(ii) Moreover, the cross-attention mechanism is adopted between the encoder and the decoder to learn the CSI-beamforming  dependence, 
thus refining the transmit bemaforming and pinching beamforming coefficients in correspondence to CSI conditions.
The detailed designs can be illustrated in the sequel.

\subsubsection{Encoder Design}
The self-attention mechanism of Transformer allows each element (which is also known as \textit{token}) in the sequence to 
interact with other elements to model their global dependencies. 
Given the vectorized CSI 
input $\mathbf{z} \!= \! \big\{\mathbf{x}^{\mathrm{U}}, \mathbf{y}^{\mathrm{U}}\big\} \!\in\! \mathbb{R}^{2K\times 1}$, 
we first embed $\mathbf{z}$ into feature $\mathbf{Z}^{\mathrm{E}} \in \mathbb{R}^{2K \times N_{\mathrm{model}}}$ by sinusoidal position embedding: 
\begin{equation}\label{input_embedding}
\mathbf{Z}^{\mathrm{E}}  = \mathbf{w}^{\mathrm{E}} \mathbf{z}^{\top} + \mathbf{PE},
\end{equation}
where $N_{\mathrm{model}}$ denotes the feature embedding dimension, and $\mathbf{w}^{\mathrm{E}}\in\mathbb{R}^{N_{\mathrm{model}}\times 1}$ is a learnable linear transform. 
$\mathbf{PE}$ represents the fixed sinusoidal position embedding that retains the sequential order of each token, 
which is defined as
$\mathbf{PE}(i,2j) = \sin\left(\frac{i}{10000^{2j/N{\mathrm{model}}}}\right)$,
$\mathbf{PE}(i,2j+1) = \cos\left(\frac{i}{10000^{2j/N{\mathrm{model}}}}\right)$, 
$j=0,1,\dots,N_{\mathrm{model}}/2-1$.

The encoder exploits a $J_{\mathrm{E}}$-layer structure. 
In each encoding layer, given the embedding feature (i.e., hidden state) $\mathbf{Z}^{\mathrm{E}}$, 
Transformer generates three different matrices, 
named \textit{query} $\mathbf{Y}_{i}^{\mathrm{Q}}$, \textit{key} $\mathbf{Y}_{j}^{\mathrm{K}}$, and \textit{value} $\mathbf{Y}_{j}^{\mathrm{V}}$. 
Specifically, \textit{query} $\mathbf{Y}_{i}^{\mathrm{Q}}$ indicates which information requires attentions by each element,  
and \textit{key} $\mathbf{Y}_{j}^{\mathrm{K}}$ signifies what information can be provided by each element. 
Hence, the self-attention score $\mathbf{Y}^{\mathrm{Q}}\left(\mathbf{Y}^{\mathrm{K}}\right)^\top$, defined as the inner product of \textit{query} and \textit{key}, 
denotes the correlation between elements. 
Moreover, \textit{value} $\mathbf{Y}_{j}^{\mathrm{V}}$ represents the actual information maintained by each element, which will be multiplied with the self-attention score 
for hidden state update. 
We employ multiple attention heads to capture different inter-dependence patterns. 
Each attention head $i$ individually computes \textit{query}, \textit{key}, and \textit{value} matrices by 
\begin{equation}\label{linear_projection}
    \mathbf{Y}_{i}^{\mathrm{K}} \!=\! \mathbf{Z}^{\mathrm{E}} \mathbf{W}_{i}^{\mathrm{K}},\quad  
    \mathbf{Y}_{i}^{\mathrm{Q}} \!=\! \mathbf{Z}^{\mathrm{E}} \mathbf{W}_{i}^{\mathrm{Q}}, 
    \quad \mathbf{Y}_{i}^{\mathrm{V}} \!=\! \mathbf{Z}^{\mathrm{E}} \mathbf{W}_{i}^{\mathrm{V}},
\end{equation}
where $I$ denotes the total number of attention heads, 
$\mathbf{W}_{i}^Q, \mathbf{W}_{i}^K, \mathbf{W}_{i}^V \in \mathbb{R}^{N_{\text{model}} \times N_{\text{model}}}$ denotes the learnable weight matrices. 
% The self-attention score is defined as the inner product of the query matrix $\mathbf{Y}_{i}^{\mathrm{Q}}$ and the key matrix $\mathbf{Y}_{i}^{\mathrm{K}}$, which captures the correlations of elements. 
By normalizing the self-attention scores using the Softmax function and multiplying it with \textit{value} matrix, each head $i$ obtains attention scores as follows:
\begin{equation}\label{self_attention}
    \text{head}_{i} \!\!=\!\! Attn(\mathbf{Y}_{i}^{\mathrm{Q}}, \mathbf{Y}_i^{\mathrm{K}}, \mathbf{Y}_i^{\mathrm{V}}) 
    \!\!=\!\! \text{softmax}\!\!\left(\tfrac{\mathbf{Y}_{i}^{\mathrm{Q}}\!\left(\mathbf{Y}_{i}^{\mathrm{K}}\right)^\top}{\sqrt{N_{\mathrm{model}}}}\!\right)\!\mathbf{Y}_{i}^{\mathrm{V}}\!.
\end{equation}
Then, the encoder concatenates attention weights from different heads, 
and computes multi-head features through a learnable matrix $\mathbf{W}^{O}$ and a feed-forward network (FFN) $\mathcal{W}_{\mathrm{FFN}}$. 
% \begin{equation}\label{Encoder_Output}
%     % \begin{split}
%        % &\text{MultiHead}\left(\left\{\mathbf{Y}_{i}^{\mathrm{Q}}\right\}, 
%         % \left\{\mathbf{Y}_{i}^{\mathrm{K}}\right\}, 
%         % \left\{\mathbf{Y}_{i}^{\mathrm{V}}\right\}\right) = % \\ &
%         % \quad \quad 
%         \text{multihead} = \mathrm{FFN}\left(\text{Concat}(\text{head}_1, \text{head}_2, ..., \text{head}_I)\mathbf{W}^{O}\right).
%     % \end{split}
% \end{equation}
The hidden state of each encoding layer is updated by the multi-head features using a residual connection layer:
\begin{equation}\label{Encoder_Output}
    \mathbf{Z}^{\mathrm{E}}\! \leftarrow\! \mathbf{Z}^{\mathrm{E}} \!+\! 
    % \text{MultiHead}\left(\left\{\mathbf{Y}_{i}^{\mathrm{Q}}\right\},\left\{\mathbf{Y}_{i}^{\mathrm{K}}\right\}, 
    % \left\{\mathbf{Y}_{i}^{\mathrm{V}}\right\}\right) ,
    \mathcal{W}_{\mathrm{FFN}}\left(\text{Concat}(\text{head}_1, \text{head}_2, ..., \text{head}_I)\mathbf{W}^{O}\right),
\end{equation}
which is further used as the input of the next encoding layer. 
\subsubsection{Decoder Design}
The decoder dynamically attends to relevant information in the final hidden state $\mathbf{Z}^{\mathrm{E}}$ obtained from the encoder, and predicts the primal/dual variables via a cross-attention mechanism.
Let $\mathbf{Z}^{\mathrm{D}} \in \mathbb{R}^{S_{\mathrm{dec}} \times N_{\mathrm{model}}}$ denote the sequence of decoder prediction embeddings, where $S_{\mathrm{dec}}$ is the output sequence length.
Specifically, the initial decoder input is constructed as
\begin{equation}\label{Decoder_initialization}
\mathbf{Z}^{\mathrm{D}} = \mathbf{Z}_{\mathrm{init}}^{\mathrm{D}} + \mathbf{PE},
\end{equation}
where $\mathbf{Z}_{\mathrm{init}}^{\mathrm{D}}$ is a learnable parameter matrix and 
$\mathbf{PE}$ is the sinusoidal positional encoding applied to each decoder element.
At each decoding layer, each attention head $i$ of the decoder generates the key matrix for $\mathbf{Z}^{\mathrm{D}}$, 
while computing query matrix and value matrix using the encoder state $\mathbf{Z}^{\mathrm{E}}$:
\begin{equation}\label{linear_projection_decoder}
    \widetilde{\mathbf{Y}}_{i}^{\mathrm{K}} \!=\! \mathbf{Z}^{\mathrm{D}} \widetilde{\mathbf{W}}_{i}^{\mathrm{K}},\quad  
    \widetilde{\mathbf{Y}}_{i}^{\mathrm{Q}} \!=\! \mathbf{Z}^{\mathrm{E}} \widetilde{\mathbf{W}}_{i}^{\mathrm{Q}}, 
    \quad\widetilde{\mathbf{Y}}_{i}^{\mathrm{V}} \!=\! \mathbf{Z}^{\mathrm{E}} \widetilde{\mathbf{W}}_{i}^{\mathrm{V}},
\end{equation}
where $\widetilde{\mathbf{W}}_{i}^{\mathrm{K}}$, $\widetilde{\mathbf{W}}_{i}^{\mathrm{Q}}$, $\widetilde{\mathbf{W}}_{i}^{\mathrm{V}}$ are trainable decoder parameters. 
The cross attention score between $\mathbf{Z}^{\mathrm{E}}$ and $\mathbf{Z}^{\mathrm{D}}$ is obtained by:
\begin{equation}\label{crosb_attention}
    \mathrm{head}_{i} \!\!=\!\! CrAttn\!\!\left(\widetilde{\mathbf{Y}}_{i}^{\mathrm{K}},\!\widetilde{\mathbf{Y}}_{i}^{\mathrm{Q}},\!\widetilde{\mathbf{Y}}_{i}^{\mathrm{V}}\right)
    \!=\! \text{softmax}\!\!\left(\!\tfrac{\widetilde{\mathbf{Y}}_{i}^{\mathrm{Q}}\!
    \left(\widetilde{\mathbf{Y}}_{i}^{\mathrm{K}}\right)^\top}{\sqrt{N_{\mathrm{model}}}}\!\right)\!
    \widetilde{\mathbf{Y}}_{i}^{\mathrm{V}}\!.
\end{equation}
Similar to the encoder, the output is predicted by 
\begin{equation}\label{Decoder_Output}
    \mathbf{Z}^{\mathrm{D}} \leftarrow \mathbf{Z}^{\mathrm{D}} + \text{MultiHead} \left(\widetilde{\mathbf{Y}}_{i}^{\mathrm{K}} , \widetilde{\mathbf{Y}}_{i}^{\mathrm{Q}}, 
    \widetilde{\mathbf{Y}}_{i}^{\mathrm{V}}\right).
\end{equation}

\begin{algorithm}[!t]
    \caption{The Proposed KDL-Transformer Algorithm}\label{alg:transformer}
    \textbf{Input}: Vectorized user location (CSI) sequence $\mathbf{z}\in \mathbb{R}^{2K \times 1}$. \\
    \textbf{Output}: Transmit beamforming and pinching beamforming. \\
    \textbf{Parameters}: Embedding dimension $N_{\text{model}}$, number of heads $I$, numbers of encoder and decoder layers $J_{\mathrm{E}}$ and $J_{\mathrm{D}}$.
    
    \begin{algorithmic}[1] 
      \STATE Compute feature embedding $\mathbf{Z}^{\mathrm{E}}$ by \eqref{input_embedding}.
      \FOR{each encoding layer in $\mathcal{E}$}
      \STATE Compute query $\mathbf{Y}_{i}^{\mathrm{Q}}$, key $\mathbf{Y}_{i}^{\mathrm{K}}$, and value $\mathbf{Y}_{i}^{\mathrm{V}}$ for each attention head $i$ by \eqref{linear_projection}. 
      \STATE Obtain self-attention by \eqref{self_attention} and update $\mathbf{Z}^{\mathrm{E}}$ by \eqref{Encoder_Output}.
      \ENDFOR
      \STATE Initialize decoder output $\mathbf{Z}^{\mathrm{D}}$ by \eqref{Decoder_initialization}.
      \FOR{each decoding layer in $\mathcal{D}$}
      \STATE Calculate cross attention by \eqref{linear_projection_decoder} and \eqref{crosb_attention}.
      \STATE Update $\mathbf{Z}^{\mathrm{D}}$ by \eqref{Decoder_Output}.
      \ENDFOR
      \STATE Obtain PA deployment $\mathbf{x}^{\mathrm{end}}\in\mathbb{R}^{N\times 1}$, $\bm{\omega}\in\mathbb{R}^{N\times L}$, 
    dual variable $\bm{\lambda}\in\mathbb{R}^{K\times 1}$, and power allocation $\bm{\mu}\in\mathbb{R}^{K\times 1}$ from $\mathbf{Z}^{\mathrm{D}}= 
    \big[\mathbf{x}^{\mathrm{end}}, \bm{\omega}, \bm{\lambda}, \bm{\mu}\big] \!\in \mathbb{R}^{ (N+M+2K)\times 1}$. 
      \STATE Project $\mathbf{x}^{\mathrm{end}}$, $\bm{\omega}$ by \eqref{project_x_end}, \eqref{project_spacing} and obtain $\mathbf{X}$. 
      \STATE Construct KKT-conditioned solution $\mathbf{D}$ by \eqref{DBF_KKT_matrix} and \eqref{project_DBF}.
      \RETURN $\mathbf{X}$, $\mathbf{D}$.
    \end{algorithmic}
\end{algorithm}

\subsubsection{Constraint Guarantees}
We introduce projection operations in this part to strictly ensure constraints (C1) - (C3). 
Specfically, the transformer decoder predicts primal and dual variables $\left\{\mathbf{x}^{\mathrm{end}}, \bm{\omega}, \bm{\mu}, \bm{\lambda}\right\}$, 
and reconstruct feasible solutions of $\mathbf{X}$ and $\mathbf{D}$ through the introduced projections.  
Vector $\mathbf{x}^{\mathrm{end}} = \big[x_{1}^{\mathrm{end}}, x_{2}^{\mathrm{end}}, \dots, x_{N}^{\mathrm{end}}\big]^{\top}\in\mathbb{R}^{N\times 1}$ 
assigns the last PA's position along each waveguide. 
Moreover, $\bm{\omega}   = \left[\omega_{n,l}\right]\in\mathbb{R}^{N\times L}$ denotes the spacing between any two PAs along the $n$-th waveguide, 
where $\omega_{n,1}$ is the distance from the first PA to the feed point of waveguide $n$, 
and $\omega_{n,l} = x_{n,l} - x_{n,l-1}$, $\forall l > 1$, indicates the mutual intervals between the remaining PAs. 

To satisfy PAs' range limitation (C2), we first project $x_{n}^{\mathrm{end}}$ into the feasible region 
$L \Delta_{\min} \leqslant x_{n}^{\mathrm{end}} \leqslant S_{\mathrm{x}}$ using 
\begin{equation}\label{project_x_end}
    x_{n}^{\mathrm{end}} \!\leftarrow\! L\Delta_{\min} \!+\! \mathrm{Sigmoid}\!\left(x_{n}^{\mathrm{end}}\right)\!\left(S_{\mathrm{x}}\!-\!L\Delta_{\min}\right)\!, ~
    \forall n \!\in\! \mathcal{N},
\end{equation}
where $\mathrm{Sigmoid}(\cdot)$ is the Sigmoid activation function. 
Further, to simultaneously ensure the minimum spacing constraint (C1), the following projection operation is introduced. 
Given a threshold $0 < \!\epsilon\! \leqslant 1/L$ and a predicted vector $\mathbf{z} = \left[z_{1}, z_{2}, \dots, z_{L}\right]^{\top}$, $0 \!\leqslant\! z_{l} \!\leqslant\! 1$, 
the projection operation 
\begin{equation}
        \mathbf{z}_{\mathrm{proj}} 
        = \mathcal{P}\left(\mathbf{z}, \epsilon\right) 
        = \epsilon \mathbf{1} + \frac{1-L\epsilon}{\sum_{l=1}^{L}z_{l}}\mathbf{z},
\end{equation}
ensures that $z_{\mathrm{proj},l} \!\geqslant\! \epsilon$, $\forall l$, and $\sum\limits_{l=1}z_{\mathrm{proj},l} \!=\! 1$.
Define $\bm{\omega}_{n}^{T} \!\triangleq\! \big[\omega_{n,1}, \allowbreak \omega_{n,2}, \dots, \omega_{n,L}\big]\!\in\!\mathbb{R}^{1\times L}$. 
We enforce the minimum spacing constraints (C1) by applying projection $\mathcal{P}\left(\cdot\right)$ on $\bm{\omega}_{n}$:
\begin{equation}\label{project_spacing}
    \bm{\omega}_{n} \leftarrow x_{n}^{\mathrm{end}} ~ \mathcal{P}\left(\bm{\omega}_{n},\tfrac{\Delta_{\min}}{x_{n}^{\mathrm{end}}}\right), ~\forall n\in\mathcal{N}.
\end{equation}
As a result, the $l$-th PA's position $x_{n,l}$ on waveguide $n$ can be obtained by accumulating the intervals of previous $l$ PAs using
$x_{n,l} = \sum_{i = 1}^{l} \omega_{n,i}, ~ \forall l \in \mathcal{L}, ~\forall n \in \mathcal{N}$.

Moreover, we ensure the transmitting power constraint (C3) by projecting $\mu_{k}$ into range $\left[0, P\right]$ and normalizing $\mathbf{D}$ using 
\begin{equation}\label{project_DBF}
    \mu_{k} \leftarrow \frac{\mu_{k}P}{\sum_{i\in\mathcal{K}}\mu_{i}}, 
    \quad 
    \mathbf{D}^{\mathrm{KKT}} \leftarrow 
    \frac{\mathbf{D}^{\mathrm{KKT}}\text{diag}\left(\bm{\mu}^{1/2}\right)}{\sum_{i\in\mathcal{K}}\left\Vert\mathbf{d}_{i}^{\mathrm{KKT}}\right\Vert^{2}}.
\end{equation} 

Algorithm \ref{alg:transformer} summarizes the computation procedure for the developed KDL-Transformer algorithm for predicting solutions.
The encoder and the decoder are jointly trained through end-to-end learning to minimize the loss function, i.e., the negative system sum rate. 
The inference time complexity can be analyzed as follows. 
The encoder processes user locations with a sequence length of $S_{\mathrm{enc}} \!=\! 2K$, 
while the decoder outputs $\big[\boldsymbol{x}_{\mathrm{end}}, \boldsymbol{\omega}, \boldsymbol{\lambda}, \boldsymbol{\mu}\big]$ 
with a sequence length of $S_{\mathrm{dec}} \!=\! N \!+\! NL \!+\! 2K$. 
Given the embedding width $N_{\mathrm{model}}$ and $J_E$ ($J_D$) encoder (decoder) layers, 
the forward propagation of the KDL-Transformer includes: 
(i) encoder self-attention and feed-forward with complexity $\mathcal{O}\!\big(J_E(S_{\mathrm{enc}}^2 N_{\mathrm{model}} \!+\! S_{\mathrm{enc}} N_{\mathrm{model}}^2)\big)$; 
(ii) decoder self-attention, cross-attention, and feed-forward 
with $\mathcal{O}\!\big(J_D(S_{\mathrm{dec}}^2 N_{\mathrm{model}} \!+\! S_{\mathrm{dec}} N_{\mathrm{model}}^2 \!+
    \! S_{\mathrm{dec}} S_{\mathrm{enc}} N_{\mathrm{model}})\big)$; 
and (iii) KKT-conditioned beamformer reconstruction in \eqref{DBF_KKT_matrix}, requiring $\mathcal{O}(N^3)$. 
The overall inference complexity is thus 
    $\mathcal{O}\!\big((J_E S_{\mathrm{enc}}^2 \!+\! J_D S_{\mathrm{dec}}^2 \!+\! 
    J_D S_{\mathrm{dec}} S_{\mathrm{enc}}) N_{\mathrm{model}} \!+\! (J_E S_{\mathrm{enc}} \!+\! J_D S_{\mathrm{dec}}) 
    N_{\mathrm{model}}^2 \!+\! N^3\big)$.

\begin{remark}[Complexity Comparsions]
Considering $L > N = K$ and $S_{\mathrm{dec}} = N + NL + 2K$, the inference complexity of KDL-Transformer 
mainly scales as $\mathcal{O}\big(J_D N_{\mathrm{model}} N^2 L^2\big)$. 
Compared with MM-PDD, whose time complexity is dominated by 
$\mathcal{O}\big(N_{\mathrm{it}} N^3 L^3\big)$ (typically with $N_{\mathrm{it}} > 50$), 
the KDL-Transformer achieves a much lower cost by fixing $J_D$ and $N_{\mathrm{model}}$, 
such that $J_D \ll N_{\mathrm{it}}$ and $N_{\mathrm{model}} \leqslant N^{1.5} L^{1.5}$. 
\end{remark}

\begin{table*}[!b]
    \centering    \caption{Detailed comparisons of numerical results over NVIDIA A40 workstation.}
    \resizebox{1\textwidth}{!}{ 
        \begin{tabular}{c|ccccccc}
            \toprule
            \multirow{2}{*}{\diagbox{Performance}{Method}} & MM-PDD & KDL-Transformer & KDL-Transformer-OCA & KDL-ResNet & Black-box Transformer & Black-box ResNet & Conventional MIMO\\
            &($\rho=0.01$)&(with cross attn.)&(without cross attn.)&($50$ layers) &(without cross attn.)&($50$ layers) &(WMMSE-PDD)\\\bottomrule
            Sum rate ($K=4, L=8$) & 49.76 & \textbf{65.83} & \underline{55.61} &  43.03& 15.99& 16.46& 32.14 \\ \hline
            Execution time (second) & 255.51/sample &   0.0324/batch&  0.0277/batch & \underline{ 0.0172/batch}&  0.0196/batch&  \textbf{0.0157/batch}& 157.21/sample\\ \bottomrule
            Sum rate ($K=4, L=16$) & 47.20 & \textbf{67.68}& \underline{53.01}& 45.58& 16.71& 16.75& 28.54\\\hline
            Execution time (second) & 408.12/sample & 0.0477/batch&  0.0436/batch&   0.0201/batch&  \textbf{0.0198/batch}&  \textbf{0.0198/batch}& 179.24 \\ 
            \bottomrule
        \end{tabular}\label{table_result}
    }
\end{table*}

\section{Simulation Result}
In this section, we provide numerical results to demonstrate the effectiveness of the proposed PASS design and algorithms, 
where a BS serves $K = \left\{4, 6\right\}$ single-antenna users in downlink MIMO communications. 
The system operates at $f= 30$ GHz frequency band, and the wavelength is $\lambda_{f} = 0.01$ meters.
Users are randomly distributed in a range of $S_{\mathrm{x}} \times S_{\mathrm{y}} = 20\times 10~\text{m}^{2}$ area. 
Both users and PASS have fixed heights, where $z_{k}^{\mathrm{U}}=0$, $\forall k \in\mathcal{K}$, and $h_{\mathrm{PA}}=2.5$ meter, $\forall n \in \mathcal{N}$.
To ensure multiplexing gains, the number of RF chains used by PASS is set as $N=K$. 
Each RF chain is connected to a waveguide, and each waveguide has $L=\left\{8,12,16,20,24,28,32\right\}$ PAs. 
All the waveguides stretch across a span of $20$ meters along the x-axis (horizontal) direction\footnote{
    Motion latency of PAs is ignored. To reduce timing overheads for practical motion, a semi-continuous activation~\cite{PASS_tutorial} can be adopted, 
    which activates pre-installed PAs near the optimized locations via small, fast micro-adjustments.}. 
Moreover, the waveguides are uniformly spaced along the $y$-axis (vertical) directions with an equal interval of $d_{\mathrm{W}}=S_{\mathrm{y}}/N$ meter. 
The maximum transmission power is set as $P=\big\{10, 12, 14,16,18, 20, 22, 24 \big\}$ dBm, and 
the noise power is $\sigma^2 = -90$ dBm. 
The refraction index is given by $n_{\mathrm{eff}} = 1.4$. 
The residual tolerance for the termination of MM-PDD algorithm is set as $\varepsilon= 10^{-6}$. 
For the initialization of MM-PDD algorithm, PAs are equally spaced around the average user location, $\mathbf{D}^{(0)}$ is initialized 
by regularized zero-forcing solution, dual variables $\bm{\lambda}^{(0)} = \mathbf{0}$, penalty factor $\rho^{(0)} = 10^{-4}$, and other auxiliary variables are randomly initialized.
Moreover, $J_E=J_D=2$, and $N_{\mathrm{model}}=128$.
We generate $30000$ samples by Monte-Carlo simulation to train KDL-Transformer\footnote{
    Reproducible code for KDL-Transformer is publicly available at \url{https://github.com/xiaoxiaxusummer/KDL_Transformer_Beamforming}.}. 
For each setting of $(P, L, K, S_x)$, all numerical results are averaged over 64 i.i.d. random test samples. Increasing the size of the test dataset does not noticeably affect the experimental results.
Several baseline algorithms are considered: 
\begin{itemize}
    \item \textbf{Massive MIMO}: The conventional massive MIMO array based on hybrid beamforming architecture~\cite{SubConnectedHybridBF} is adopted at the BS located at the origin point, where PDD algorithm~\cite{PDD_HybridBF} is utilized to jointly optimize the analog beamforming and the transmit beamforming. 
    Aligned with the proposed PASS, the massive MIMO uses a sub-connected structure, where each RF chain is only connected to a subset of $L$ antennas through $L$ phase shifters, 
    thus reducing the hardware costs and complexity. 
    \item \textbf{Black-box Transformer/ResNet}: The end-to-end unsupervised learning is adopted, which trains a black-box Transformer and a 50-layer ResNet for joint transmit and pinching beamforming, respectively.
    \item \textbf{KDL-ResNet}: The proposed KDL is implemented by training a 50-layer ResNet model. 
    \item \textbf{KDL-Transformer without cross attention (KDL-Transformer-OCA)}: The proposed KDL is implemented by training a self-attention Transformer that does not involve the cross-attention mechanism. 
\end{itemize}

\begin{figure}[!t]
    \centering
    \includegraphics[width=0.49\textwidth]{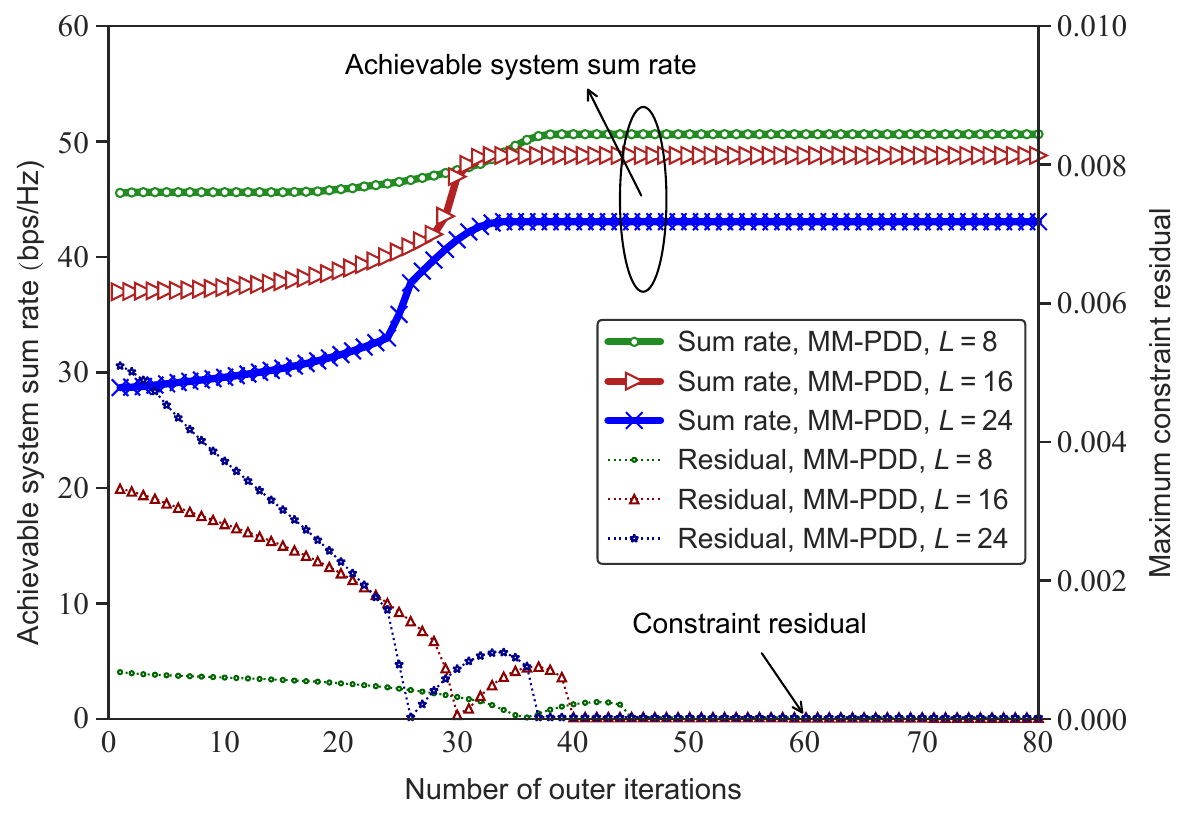}
    \caption{Convergence behaviors of the developed MM-PDD algorithm. $K=4$, $P=10$ dBm, $S_{\mathrm{x}}=20$ m. 
     }
    \label{fig_conv_PDD}
\end{figure}

\begin{figure}[!t]
    \centering
    \includegraphics[width=0.49\textwidth]{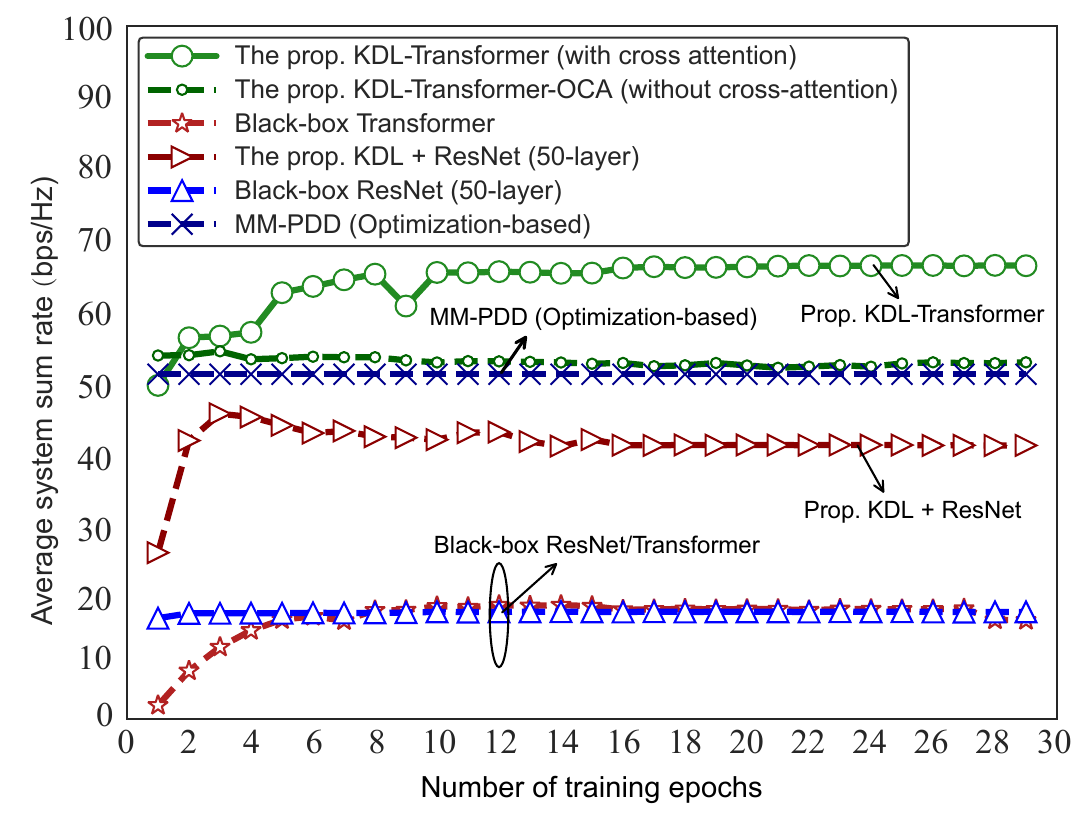}
    \caption{Comparisons of convergence performance of the proposed KDL-Transformer with other learning-based algorithms. 
    $K=4$, $L=8$, $P=10$ dBm, $S_{\mathrm{x}}=20$ m. } 
    \label{fig_conv_KDL}
\end{figure}

Fig. \ref{fig_conv_PDD} demonstrates the convergence behaviors of the developed MM-PDD algorithm in terms of both the achieveable system sum rate and the 
maximum constraint residual, i.e., $\left\Vert\mathbf{B}\right\Vert_{\infty}$. 
As shown in Fig. \ref{fig_conv_PDD}, by appropriately initializing the penalty factor $\rho$, the developed MM-PDD algorithm can successfully seek the directions to   
increase the system sum rate and reduce the constraint residuals simultaneously. 
Specifically, the maximum constraint residuals $\left\Vert\mathbf{B}\right\Vert_{\infty}$ rapidly decreases within the first $40$ iterations. 
As the punishments on constraint residuals become small, the achievable system sum rate will dramatically enhance and finally achieve convergence. 
At the convergence, the constraint residual eventually approach $0$ and becomes stable, which implies that the equality constraints can be satisfied at convergence. 
When different numbers of PAs $L$ are activated along each waveguide, 
the developed MM-PDD algorithm can reach convergence within $50$ iterations, which verifies  its efficiency. 
Furthermore, when $L$ increases, the convergence speed would not be decreased, as the punishments on equation violation increase with $L$, which will enforce 
PAs to quickly determine their locations.

\begin{figure}[!t]
    \centering
    \includegraphics[width=0.49\textwidth]{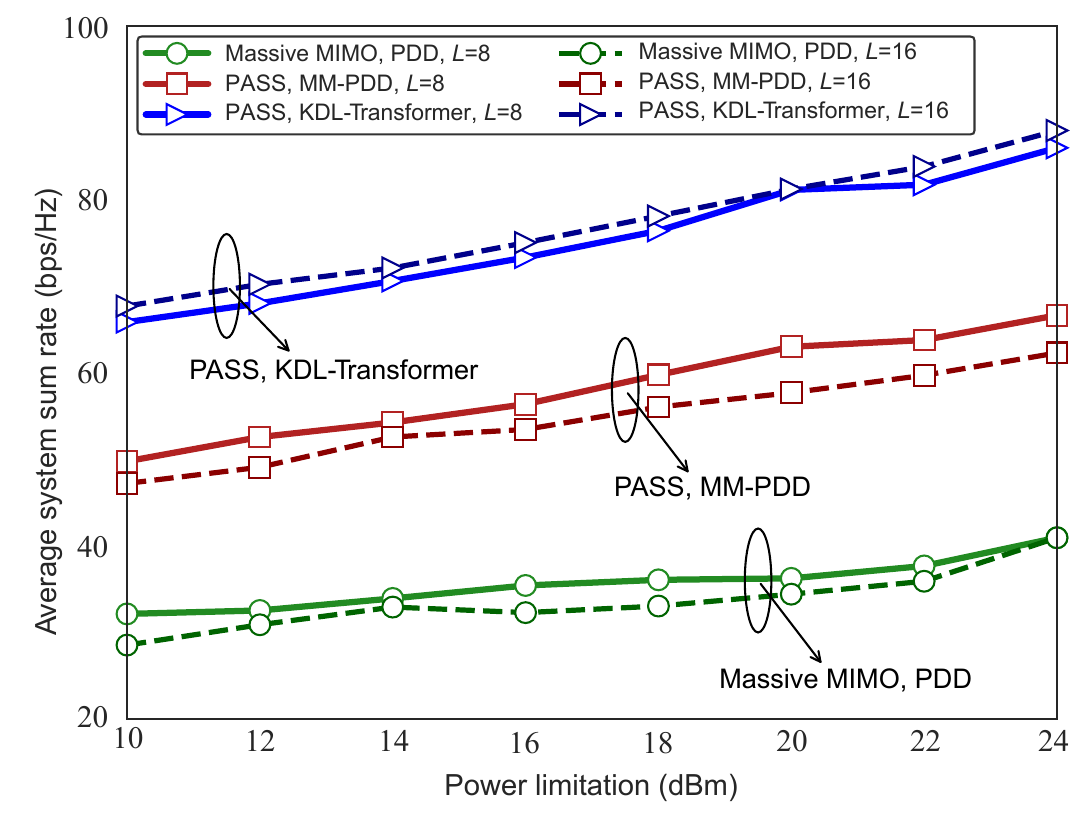}
    \caption{Performance comparisons of different algorithms under different $P$. $K=4$, $S_{\mathrm{x}}=20$ m. } 
    \label{fig_rate_P}
\end{figure}

Fig. \ref{fig_conv_KDL} exhibits the convergence behaviors of the proposed KDL-Transformer algorithm and other learning algorithms, 
where the eventual performance achieved by MM-PDD algorithm is also displayed for comparisons. 
The detailed numerical results are further presented in Table \ref{table_result}.
As shown in Fig. \ref{fig_conv_KDL}, the black-box learning algorithms based on both ResNet and Transformer 
perform much worse than the MM-PDD algorithm. 
This is because the coupled beamforming optimization problem has numerous local optimums and saddle points, 
making it difficult for the unsupervised learning 
to seek high-quality solutions by solely relying on gradient descent. 
By contrast,  despite the utilized ML models, the proposed KDL method can significantly enhance the performance 
of learning-based algorithms, 
which confirms its efficiency to approximate KKT points. 
Notably, the proposed KDL-Transformer algorithm improves over $20\%$ performance than the MM-PDD algorithm.
This demonstrates its potentials to circumvent the inefficiency of iterative alternating block descent required by the optimization algorithms. 
Furthermore, the proposed KDL-Transformer improves over $40\%$ and over $18\%$ performance then \textit{KDL-ResNet} and \textit{KDL-Transformer-OCA}, which confirms 
that its efficiency in characterizing both inter-PA/inter-user and CSI-beamforming dependencies.

Fig. \ref{fig_rate_P} shows the system performance of different algorithms under various transmitting power limitations  $P$. 
The system sum rates of both the conventional massive MIMO and the proposed PASS framework increase with $P$. 
Both the MM-PDD algorithm and the KDL-Transformer algorithm  significantly improves the system sum rate compared to the conventional massive MIMO architecture, 
which confirms the efficiency of the proposed PASS architecture in reconfiguring large-scale path loss 
and spatial multiplexing for multi-user communciations. 
Moreover, the developed KDL-Transformer algorithm achieves the highest system performance, 
and the achieved performance gain increases with $P$, 
which demonstrates its potential to enhance the resource utilization efficiency. 

\begin{figure}[!t]
    \centering
    \includegraphics[width=0.49\textwidth]{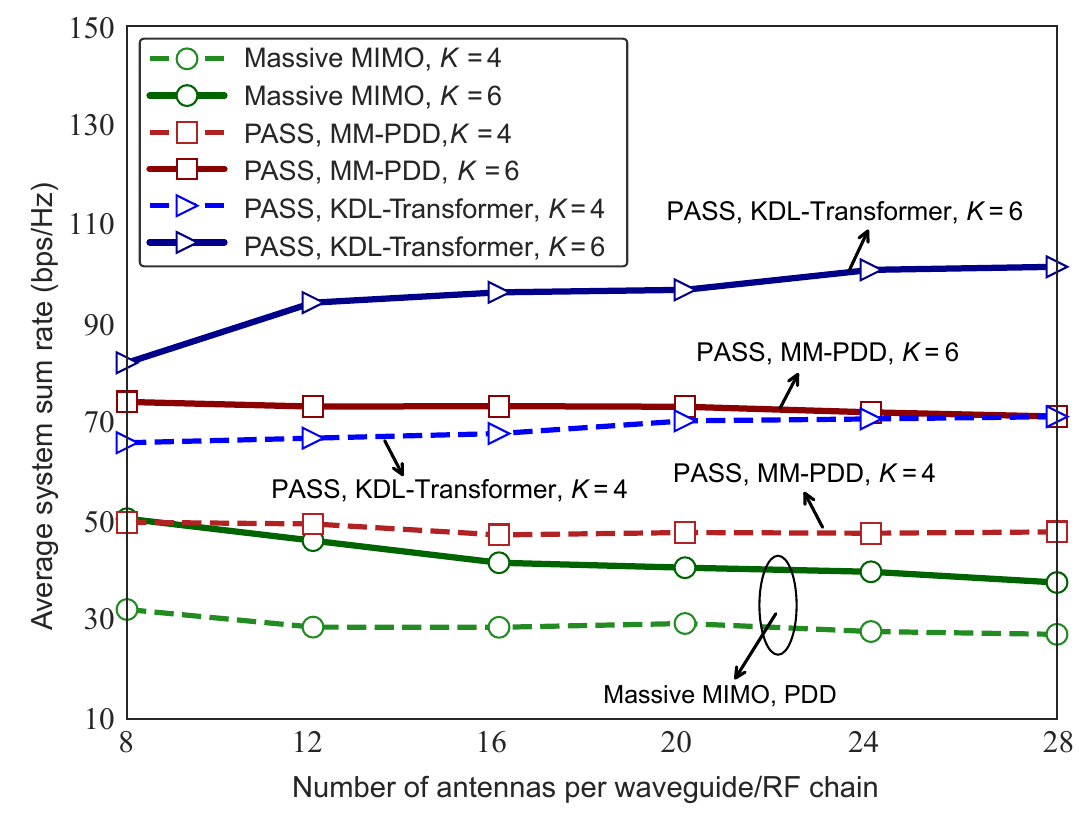} 
    \caption{Comparisons of average sum rate of the proposed algorithms under different $L$. $P = 10$ dBm, $S_{\mathrm{x}}=20$ m. } 
    \label{fig_rate_L}
\end{figure}

\begin{figure}[!t]
    \centering
    \includegraphics[width=0.49\textwidth]{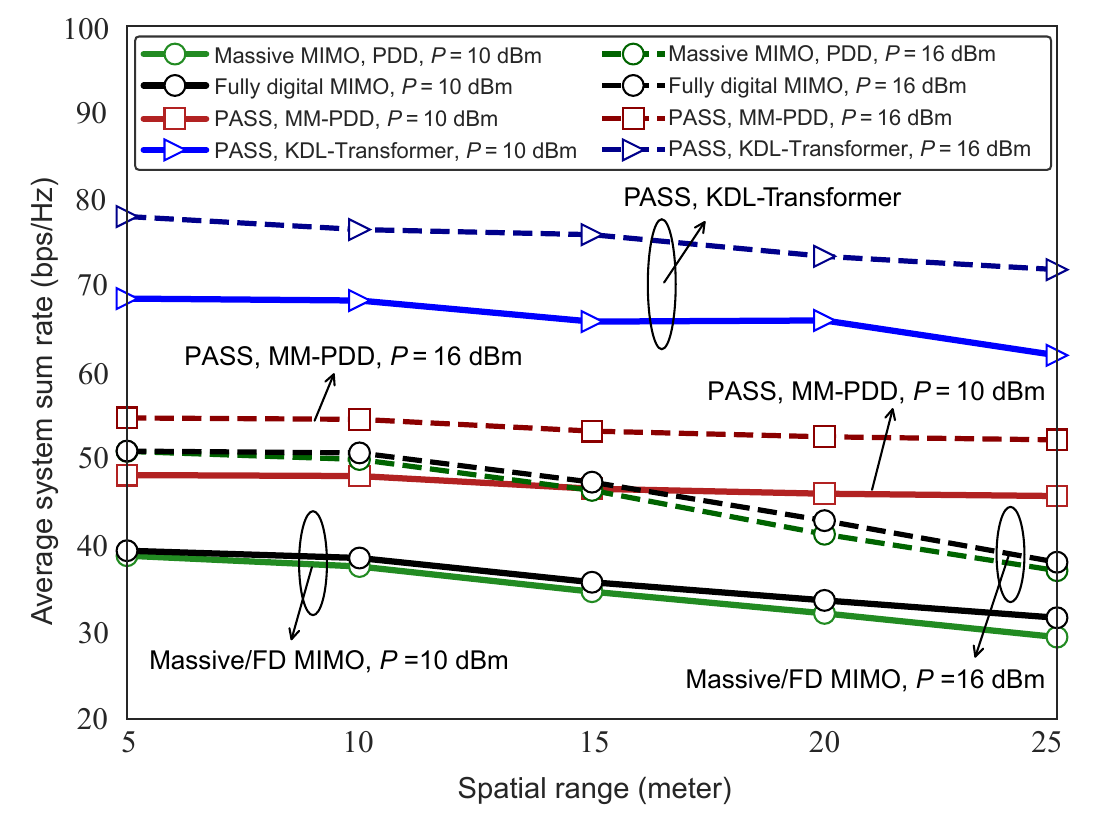} 
    \caption{Comparisons of average sum rate of the proposed algorithms under different spatial ranges $S_{\mathrm{x}}$. $K=4$, $L=8$.}
    \label{fig_rate_S}
\end{figure}

Fig. \ref{fig_rate_L} compares the system sum rates of different approaches when different numbers of pinches/antennas are activated on each waveguide/RF chain.  
Specifically,  the system sum rates of both massive MIMO and PASS increase with the number of associated users $K$, which confirms their effectivity in spatial multiplexing.  
Nevertheless, as the number of activated antennas $L$ increases, the performance of both the massive MIMO and the MM-PDD algorithm deteriorates due to increased mutual interference.
By contrast, the performance achieved by KDL-Transformer algorithm may be improved with a larger number of activated PAs. 
This implies that the optimization-based algorithms may prematurely fall into local optimums, but the pinching beamforming solutions learnt by KDL-Transformer can explore more efficient interference suppression, thus leveraging the 
the pinching deployment agility when more PAs are activated.

Fig. \ref{fig_rate_S} further demonstrates the performance of different approaches as the spatial range $S_{\mathrm{x}}$ varies from $5$ m to $25$ m.
For comparison, an ideal fully digital (FD) MIMO system is also included as an additional baseline, where $NL$ antennas are each connected to an RF chain to serve $K$ users, and fully digital beamforming is obtained via the WMMSE algorithm.
It can be observed that massive MIMO systems, with the proposed PDD-based hybrid beamforming, can closely approach the average sum rate of FD-MIMO.
However, the performance of both massive MIMO and FD-MIMO systems degrades as the spatial range increases, especially under limited transmit power.
In contrast, the performance of PASS is slightly affected as the spatial range increases, which demonstrates PASS's unique capability to adjust large-scale path loss.
Furthermore, although the KDL-Transformer outperforms MM-PDD on average, it is less robust across varying spatial ranges. 
This is likely because a larger spatial range leads to a sparser PA distribution and reduced array gain using KDL-Transformer.

\section{Conclusion} 
A novel PASS-enabled multi-user MISO framework has been proposed, which enables pinching beamforming to adjust both large-scale path loss and phases of radiated signals. 
A joint transmit and pinching beamforming optimization problem was formulated to maximize the system sum rate. 
To tackle this highly coupled and nonconvex problem, 
both optimization-based and learning-based methods have been proposed. 
The optimization-based method, named MM-PDD algorithm, used a convex surrogate function to handle complex exponential components and invoked PDD for problem decoupling, which was guaranteed to obtain stationary solutions. 
The learning-based method, termed KDL-Transformer, directly reconstructed KKT-conditioned solutions in a low-complexity and data-driven way, 
where the underlying inter-PA/inter-user and CSI-beamforming dependencies were effectively learned by Transformer. 
Numerical results have been provided to verify the efficiency of both the proposed PASS framework and algorithms.  
In future, the design of joint transmit and pinching beamforming under practical LoS blockages and waveguide attenuation can be further investigated.
Moreover, recent advances in accelerated learning, such as multithreading and federated reinforcement learning~\cite{UAV_planning_RIS, FederatedDRL}, can further reduce training overhead and improve real-time adaptability in PASS systems.
Prioritized information bottleneck frameworks for distributed online learning~\cite{PrioritizedInformationBottleneck} also support low-latency and adaptive decision making.

% \vspace{-1em}
\appendix
\subsection{Proof of Lemma \ref{lemma:WMMSE}}\label{proof:WMMSE}
The WMMSE problem w.r.t. $v_{k}$ can be written as 
\begin{equation}
    \min_{\left\{v_{k}\right\}} \sum_{k} \alpha_{k} J_{k}\left|v_{k}\right|^{2}
    \!-\!2\alpha_{k}\mathrm{Re}\left\{v_{k}\mathbf{h}_{k}^{H}\left(\mathbf{X}\right)\mathbf{G}\left(\mathbf{X}\right)\mathbf{d}_{k}\right\}. 
\end{equation}
Using the first-order optimality, the equalizer of MMSE receiver $v_{k}^{opt}$ can be obtained by \eqref{MMSE_detection}.
Substituting \eqref{MMSE_detection} into \eqref{MSE}, the MSE $e_{k}^{\mathrm{MSE}}$ is rewritten as
\begin{equation}\label{MSE_opt}
    e_{k}^{\mathrm{MSE}} \!=\!
    \left(1\!-\!\mathbf{d}_{k}^{H}\!\mathbf{G}_{k}^{H}\!\left(\mathbf{X}\right)\!
    \mathbf{h}_{k}\!\left(\mathbf{X}\right)\!J_{k}^{-1}
    \mathbf{h}_{k}^{H}\!\left(\mathbf{X}\right)\!\mathbf{G}\!\left(\mathbf{X}\right)\!\mathbf{d}_{k}\right).
\end{equation}
Then, the objective function in (P1) becomes convex with respect to $\alpha_{k}$. 
By fixing the remaining variables, we have 
\begin{equation}
    \alpha_{k}^{opt}\!\!=\!\!\left(1\!-\!
    \mathbf{d}_{k}^{H}\!\mathbf{G}^{H}\!\!\left(\mathbf{X}\right)\!\mathbf{h}_{k}\!\left(\mathbf{X}\right)\!
    J_{k}^{-1}
    \mathbf{h}^{H}\!\!\left(\mathbf{X}\right)\!\mathbf{G}\!\left(\mathbf{X}\right)\!\mathbf{d}_{k}\right)^{-1}\!\!
    \!\!\!=\!\left(e_{k}^{\mathrm{MSE}}\right)^{-1}\!\!\!\!.
\end{equation}
Substituting $\alpha_{k}^{opt}$ and $v_{k}^{opt}$ into \eqref{P1_obj}, problem (P1) becomes 
\begin{equation}\label{P1_equiv}
    \max_{\mathbf{x},\mathbf{D}} ~ \sum_{k=1}^{K}\left(\log_{2}\left(e_{k}^{\mathrm{MSE}}\right)^{-1}\right), 
    \quad \text{s.t.} ~ \text{(C1) - (C3)}.
\end{equation}
Combining $\left(e_{k}^{\mathrm{MSE}}\right)^{-1}\!=\!{E_{k}}/\left({\sum_{k'\ne k}I_{kk'}+\sigma^{2}}\right)\!=\!1+SINR_{k}$, 
\eqref{P1_equiv} is equivalent to (P0), which ends the proof. 

\subsection{Proof of Lemma \ref{lemma:Lipschitz}}\label{proof:Lipschitz}
The complex number $c$ can be written as $c=|c|e^{-i\phi_{c}}$. 
Using Euler formula $e^{i\theta}=\cos(\theta)+i\sin(\theta)$, we can rewrite function $-\mathrm{Re}\left(ce^{i\left(a\theta\right)}\right)$ 
into
\begin{equation}
f\!\left(\theta\right)\!=\!
-\!\mathrm{Re}\!\left\{ce^{i\left(a\theta\right)}\right\}
\!\!=\!\!-|c|\mathrm{Re}\!\left\{e^{i\left(a\theta-\phi_{c}\right)}\right\}
\!\!=\!\!-|c|\cos(a\theta-\phi_{c}).
\end{equation} 
Hence, the first-order and the second-order derivatives of $f\left(\theta\right)$ can be given by
\begin{equation}
    \nabla_{\theta}f\!\left(\theta\right)\!=\!a|c|\sin\left(a\theta\!-\!\phi_{c}\right),
    ~
    \nabla_{\theta}^{2}f(x)\!=\!a^{2}|c|\cos\!\left(a\theta\!-\!\phi_{c}\right).
\end{equation}
Since $|\cos(a\theta-\phi_{c})|\leqslant 1$, the second-order gradient is upper bounded. 
According to the mean value theorem, the gradient of $f\left(\theta\right)$ is Lipschitz continuous, i.e., 
\begin{equation}
    \begin{split}
        \hspace{-1em}\left\Vert \nabla_{\theta}f\!\left(\theta_{1}\right)\! 
        -\! \nabla_{\theta}f\!\left(\theta_{2}\right)\! \right\Vert 
        &\leqslant \bigg(\max_{\theta\in\left[\theta_{1},\theta_{2}\right]}
        \left|\nabla_{\theta}^{2}f(\theta)\right|\bigg)\left\Vert \theta_{1} -\theta_{2} \right\Vert
        \\&\leqslant a^{2}\left|c\right| \left\Vert \theta_{1} -\theta_{2} \right\Vert, 
        ~ \forall \theta_{1}, \theta_{2} \in \mathbb{R}. 
    \end{split}
\end{equation}
Hence, the Lipschitz gradient constant is given by 
$\varrho^{\theta}=\max_{\theta}\left|\nabla_{\theta}^{2}f(\theta)\right| =a^{2}\left|c\right|$,
which ends the proof.

\subsection{Proof of Lemma~\ref{lemma:MMPDDconvergence}}\label{proof:lemma:MMPDDconvergence}
The convergence can be established by combining the descent AL updates and the PDD theory ~\cite{PDD}. 
\textbf{(i) AL objective descent.} 
Each inner iteration $t$ builds a tight MM surrogate $f_{\mathrm{MM}}(\bm{\nu};\bm{\nu}^{(t)})\!\ge\! f_{\mathrm{AL}}(\bm{\nu})$ with equality at $\bm{\nu}\!=\!\bm{\nu}^{(t)}$. 
The BCD update yields $f_{\mathrm{MM}}(\bm{\nu}^{(t+1)};\bm{\nu}^{(t)})\!\le\! f_{\mathrm{MM}}(\bm{\nu}^{(t)};\bm{\nu}^{(t)})$, hence
$f_{\mathrm{AL}}\big(\bm{\nu}^{(t+1)}\big) \!\leqslant\! 
f_{\mathrm{MM}}\big(\bm{\nu}^{(t+1)};\bm{\nu}^{(t)}\big)\!\leqslant\! 
f_{\mathrm{MM}}\big(\bm{\nu}^{(t)};\bm{\nu}^{(t)}\big)\!=\!f_{\mathrm{AL}}\big(\bm{\nu}^{(t)}\big)$ and thus 
$\{f_{\mathrm{AL}}\big(\bm{\nu}^{(t)}\big)\}$ yields a non-increasing sequence.
\textbf{(ii) Robinson condition and equality residuals.} 
The equality constraints \eqref{eq_antgain}-\eqref{eq_effgain} have block-diagonal Jacobians 
with nonzero diagonal blocks (full row rank), and inequality constraints (C1)-(C3) admit strictly feasible points. 
Thus, the sufficient condition in~\cite[Sec.~V-A]{PDD} for Robinson's condition holds for (P2). 
Let $\bm{\xi}^{(i)}\!\triangleq\!\bm{\lambda}^{(i)}\!+\!\tfrac{1}{\rho^{(i)}}\mathbf{B}^{(i)}$ at the outer interation $i$.  
If $\|\mathbf{B}^{(i)}\|_{\infty}\!\leqslant\!\varepsilon^{(i)}$, the dual update $\bm{\lambda}^{(i+1)}\!=\!\bm{\xi}^{(i)}$ gives 
$\mathbf{B}^{(i)}\!=\!\rho^{(i)}(\bm{\lambda}^{(i+1)}\!-\!\bm{\lambda}^{(i)})$, 
hence both $\|\mathbf{B}^{(i)}\|_{\infty}$ and $\|\bm{\lambda}^{(i+1)}\!-\!\bm{\lambda}^{(i)}\|_{\infty}$ 
vanish as $\varepsilon^{(i)}\!\to\!0$.  
Otherwise $\rho^{(i+1)}\!=\!\varsigma\rho^{(i)}$ with $\varsigma\!\in\!(0,1)$. If the residuals did not decrease, then 
$\|\bm{\xi}^{(i)}\|\!=\!\|\bm{\lambda}^{(i)}+\tfrac{1}{\rho^{(i)}}\mathbf{B}^{(i)}\|$ would blow up as $\rho^{(i)}\!\to\!0$, 
contradicting the boundedness of $\{\bm{\xi}^{(i)}\}$ ensured under Robinson's conditions~\cite{PDD}. 
Thus, $\|\mathbf{B}^{(i)}\|_{\infty}\!\to\!0$ vanish asymptotically.

Combining (i) and (ii), the inner-loop MM updates ensure that each AL subproblem is solved to an approximate stationary point with a diminishing approximation error, 
while the outer-loop PDD updates enforce residuals $\|\mathbf{B}^{(i)}\|_{\infty}$ to zero and keep $\{\bm{\lambda}^{(i)}\}$ bounded. 
As $i \to \infty$, the iterates $(\bm{\nu}^{(i)},\bm{\lambda}^{(i)})$ asymptotically satisfy exact KKT conditions of (P2) with vanishing approximation errors ~\cite[Theorem~3.1]{PDD}. 
Since (P2) is an equivalent reformulation of (P0), the final solution $\bm{\nu}^\star$ is a KKT point of (P0), which completes the proof.

\begin{IEEEbiography}[{\includegraphics[width=1in,height=1.25in,clip,keepaspectratio]{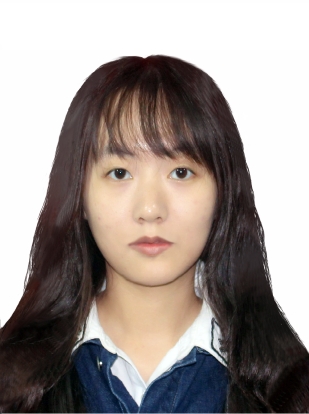}}]
    {Xiaoxia Xu} (\url{https://xiaoxiaxusummer.github.io/})
    received the B.Eng. degree and Ph.D. degree from Wuhan University,
    China, in 2017 and 2023, respectively. 
    From 2021 to 2022, she was also a visiting student at Queen Mary University of London (QMUL), U.K. Currently, she is a Postdoctoral Researcher 
    in School of Electronic Engineering and Computer Science at QMUL, London, U.K. 
    Her current research interests include millimetre-wave/terahertz communications, flexible-antenna techniques, next generation multiple access, AI for B5G/6G, and edge intelligence.
 \end{IEEEbiography}
 
 \setlength{\parskip}{0pt}
 
 \begin{IEEEbiography}[{\includegraphics[width=1in,height=1.25in,clip,keepaspectratio]{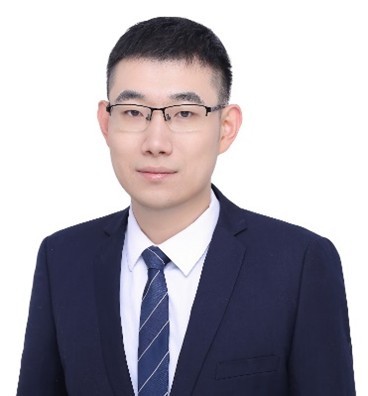}}]
    {Xidong Mu} (Member, IEEE, \url{https://xidongmu.github.io/}) received the Ph.D. degree in Information and Communication Engineering from the Beijing University of Posts and Telecommunications, Beijing, China, in 2022. He was with the School of Electronic Engineering and Computer Science, Queen Mary University of London, from 2022 to 2024, where he was a Postdoctoral Researcher. He has been a lecturer (an assistant professor) with the Centre for Wireless Innovation (CWI), School of Electronics, Electrical Engineering and Computer Science, Queen’s University Belfast, U.K. since August 2024. His research interests include flexible-antenna technologies, reconfigurable surface aided communications, next generation multiple access (NGMA), integrated sensing and communications, and optimization theory. 

    Xidong Mu is a Web of Science Highly Cited Researcher. He received the IEEE ComSoc Outstanding Young Researcher Award 
    for EMEA region in 2023 and the IEEE ComSoc Wireless Communications Technical Committee (WTC) Outstanding Young Researcher Award in 2025. He is the recipient of the 2024 IEEE Communications Society Heinrich Hertz Award, the Best Paper Award in ISWCS 2022, the 2022 IEEE SPCC-TC Best Paper Award, and the Best Student Paper Award in IEEE VTC2022-Fall. 
    He serves as the secretary of the IEEE ComSoc Technical Committee on Cognitive Networks (TCCN), the secretary of the IEEE ComSoc NGMA Emerging Technology Initiative, and the URSI UK Early Career Representative (ECR) for Commission C. He also serves as an Editor of \textsc{IEEE Transactions on Communications}, a Guest Editor for \textsc{IEEE Journal on Selected Areas in Communications}, \textsc{IEEE Transactions on Cognitive Communications and Networking}, \textsc{IEEE Internet of Things Journal}, and the ``Mobile and Wireless Networks'' symposium co-chair of IEEE GLOBECOM 2025.     
\end{IEEEbiography}

 \begin{IEEEbiography}[{\includegraphics[width=1in,height=1.25in,clip,keepaspectratio]{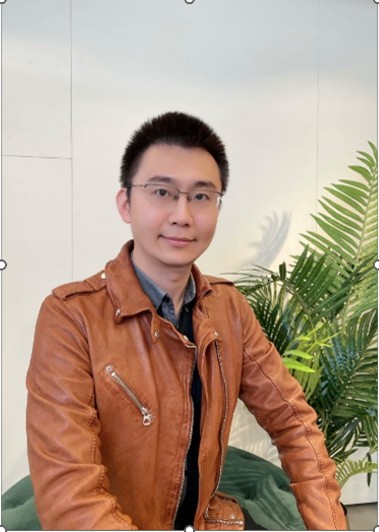}}]
     {Yuanwei Liu} (S'13-M'16-SM'19-F'24, \url{https://www.eee.hku.hk/~yuanwei/}) is a tenured full Professor in Department of Electrical and Electronic Engineering (EEE) at 
     The University of Hong Kong (HKU) and a visiting professor at Queen Mary University of London (QMUL). 
     Prior to that, he was a Senior Lecturer (Associate Professor) (2021-2024) and a Lecturer (Assistant Professor) (2017- 2021) at QMUL, London, U.K, 
     and a Postdoctoral Research Fellow (2016-2017) at King's College London (KCL), London, U.K. He received the Ph.D. degree from QMUL in 2016.  
     His research interests include non-orthogonal multiple access, reconfigurable intelligent surface, near field communications, integrated sensing and communications, and machine learning. 
 
     Yuanwei Liu is a Fellow of the IEEE, a Fellow of AAIA, a Fellow of AIIA, a Web of Science Highly Cited Researcher, an IEEE Communication Society Distinguished Lecturer, 
     an IEEE Vehicular Technology Society Distinguished Lecturer, the rapporteur of ETSI Industry Specification Group on Reconfigurable Intelligent Surfaces on work item of ``Multi-functional Reconfigurable Intelligent Surfaces (RIS): Modelling, Optimisation, and Operation'', 
     and the UK representative for the URSI Commission C on ``Radio communication Systems and Signal Processing'' (2023-2024). He was listed as one of 35 Innovators Under 35 China in 2022 by MIT Technology Review. He received IEEE ComSoc Outstanding Young Researcher Award for EMEA in 2020. He received the 2020 IEEE Signal Processing and Computing for Communications (SPCC) Technical Committee Early Achievement Award, IEEE Communication Theory Technical Committee (CTTC) 2021 Early Achievement Award. He received IEEE ComSoc Outstanding Nominee for Best Young Professionals Award in 2021. He is the co-recipient of the 2024 IEEE Communications Society Heinrich Hertz Award, the Best Student Paper Award in IEEE VTC2022-Fall, the Best Paper Award in ISWCS 2022, the 2022 IEEE SPCC-TC Best Paper Award, the 2023 IEEE ICCT Best Paper Award, and the 2023 IEEE ISAP Best Emerging Technologies Paper Award. He serves as the Co-Editor-in-Chief of IEEE ComSoc TC Newsletter, an Area Editor of IEEE Transactions on Communications and IEEE Communications Letters, an Editor of IEEE Communications Surveys \& Tutorials, IEEE Transactions on Wireless Communications, IEEE Transactions on Vehicular Technology, IEEE Transactions on Network Science and Engineering, and IEEE Transactions on Cognitive Communications and Networking. He serves as the (leading) Guest Editor for Proceedings of the IEEE on Next Generation Multiple Access, IEEE JSAC on Next Generation Multiple Access, IEEE JSTSP on Intelligent Signal Processing and Learning for Next Generation Multiple Access, and IEEE Network on Next Generation Multiple Access for 6G. He serves as the Publicity Co-Chair for IEEE VTC 2019-Fall, the Panel Co-Chair for IEEE WCNC 2024, Symposium Co-Chair for several flagship conferences such as IEEE GLOBECOM, ICC and VTC. He serves the academic Chair for the Next Generation Multiple Access Emerging Technology Initiative, vice chair of SPCC and Technical Committee on Cognitive Networks (TCCN) (2023-2024).
 \end{IEEEbiography}

 \begin{IEEEbiography}[{\includegraphics[width=1in,height=1.25in,clip,keepaspectratio]{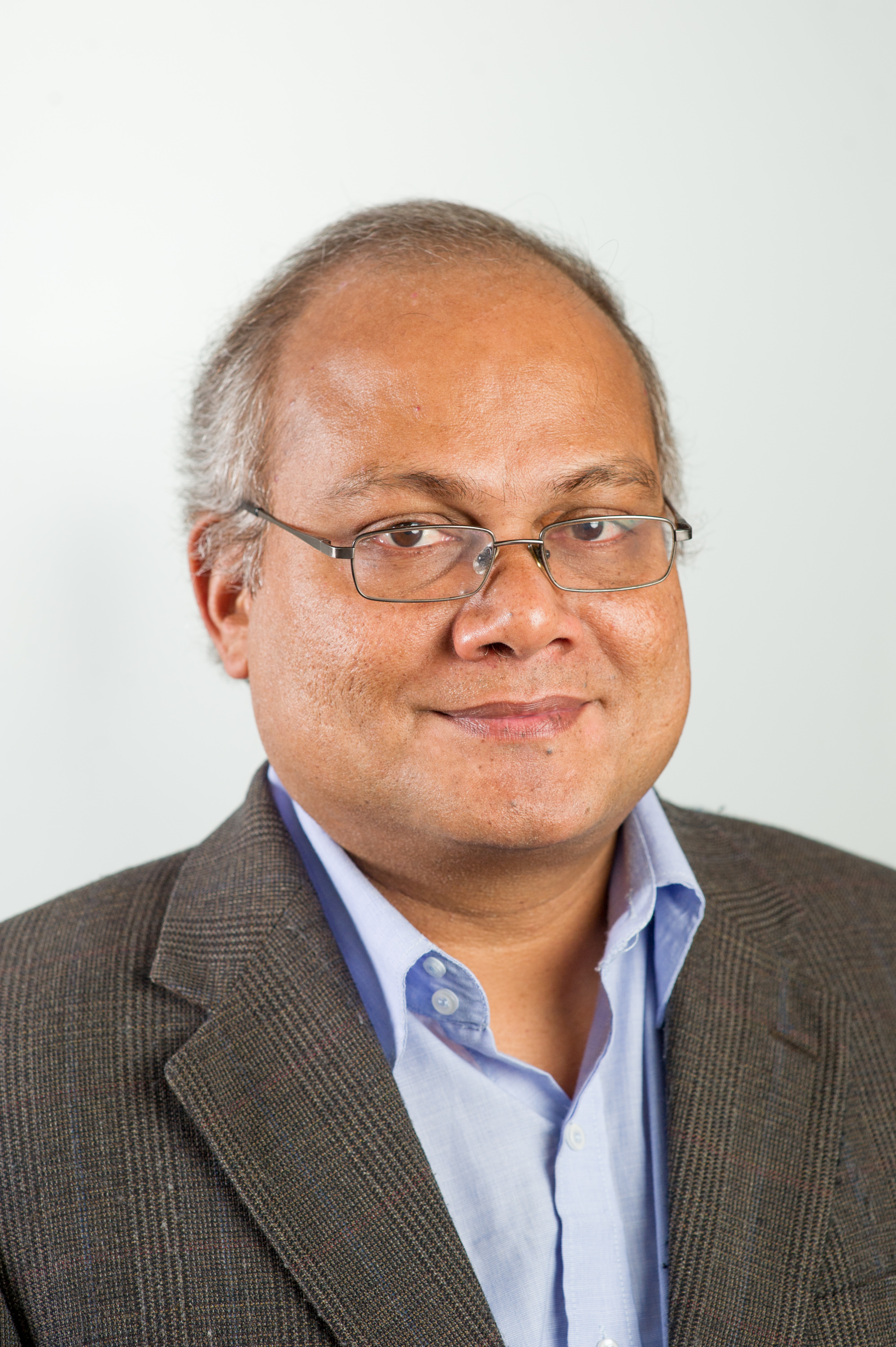}}]
     {Arumugam Nallanathan} (S'97-M'00-SM'05-F'17) is Professor of Wireless Communications and the Founding Head of the Communication Systems Research (CSR) group in the School of Electronic Engineering and Computer Science at Queen Mary University of London since September 2017. He was with the Department of Informatics at King’s College London from December 2007 to August 2017, where he was Professor of Wireless Communications from April 2013 to August 2017 and a Visiting Professor from September 2017 till August 2020. He was an Assistant Professor in the Department of Electrical and Computer Engineering, National University of Singapore from August 2000 to December 2007. His research interests include Artificial Intelligence for Wireless Systems, Beyond 5G Wireless Networks and Internet of Things (IoT). He published more than 700 technical papers in scientific journals and international conferences. He is a co-recipient of the Best Paper Awards presented at the IEEE International Conference on Communications 2016 (ICC'2016), IEEE Global Communications Conference 2017 (GLOBECOM'2017) and IEEE Vehicular Technology Conference 2018 (VTC'2018). He is also a co-recipient of IEEE Communications Society Leonard G. Abraham Prize in 2022. He is an IEEE Distinguished Lecturer. He has been selected as a Web of Science Highly Cited Researcher in 2016, and 2022-2024.  
 
 He was a Senior Editor for IEEE Wireless Communications Letters, an Editor for IEEE Transactions on Wireless Communications, IEEE Transactions on Communications, IEEE Transactions on Vehicular Technology and IEEE Signal Processing Letters. He served as a Guest Editor for numerous special issues of IEEE Journal on Selected Areas in Communications (JSAC). He served as the Chair for the Signal Processing and Communication Electronics (SPCE) Technical Committee of IEEE Communications Society and Technical Program Chair and member of Technical Program Committees in numerous IEEE conferences. He received the IEEE Communications Society SPCE outstanding service award 2012 and IEEE Communications Society RCC outstanding service award 2014.
 \end{IEEEbiography}
 

\begin{thebibliography}{1}

    \bibitem{PASS_ICCC} X. Xu, Y. Liu, X. Mu, D. Gan and A. Nallanathan, ``Beamforming for Pinching Antenna Systems (PASS): KKT-Guided Large Model Learning,'' 
    \textit{2025 IEEE/CIC International Conference on Communications in China (ICCC)}, Shanghai, China, 2025, pp. 1-6.

    \bibitem{Shannon} C. E. Shannon, ``A mathematical theory of communication,'' \textit{Bell Syst. Tech. J.}, vol. 27, no. 3, pp. 379-423, Jul. 1948. 

    \bibitem{RIS} M. Di Renzo et al., ``Smart radio environments empowered by reconfigurable intelligent surfaces: How it works, state of research, and the road
    ahead,'' \textit{IEEE J. Sel. Areas Commun.}, vol. 38, no. 11, pp. 2450-2525, 2020.

    \bibitem{STARS} X. Mu, Y. Liu, L. Guo, J. Lin and R. Schober, 
    ``Simultaneously Transmitting and Reflecting (STAR) RIS Aided Wireless Communications,'' 
    \textit{IEEE Trans. Wireless Commun.}, vol. 21, no. 5, pp. 3083-3098, May 2022.

    \bibitem{FluidAntenna} K.-K. Wong, A. Shojaeifard, K.-F. Tong, and Y. Zhang, ``Fluid antenna
    systems,'' \textit{IEEE Trans. Wireless Commun.}, vol. 20, no. 3, pp. 1950-1962, Mar. 2021.

    \bibitem{MovableAntenna} 
    L. Zhu, W. Ma, and R. Zhang, ``Modeling and performance analysis
    for movable antenna enabled wireless communications,'' \textit{IEEE Trans. Wireless Commun.}, 
    vol. 23, no. 6, pp. 6234-6250, Jun.

    \bibitem{PAr_Ding} Z. Ding, R. Schober, H. V. Poor, ``Flexible-antenna systems: A pinching-antenna perspective,'' \textit{arxiv}, arXiv:2501.10753, 2025. 

    \bibitem{PAr_Liu} Y. Liu, Z. Wang, X. Mu, C. Ouyang, X. Xu, and Z. Ding, ``Pinching Antenna Systems (PASS): Architecture
    Designs, Opportunities, and Outlook,'' \textit{arXiv preprint}, arXiv 2501.18409, 2025.

    \bibitem{PA_DOCOMO} H. O. Y. Suzuki and K. Kawai, ``Pinching antenna - using a dielectric
    waveguide as an Antenna,'' \textit{NTT DOCOMO Technical Journal}, vol. 23,
    no. 3, pp. 5-12, Jan. 2022.

    \bibitem{Rate_DL_PAr_SU} Y. Xu, Z. Ding, G. K. Karagiannidis, ``Rate Maximization for Downlink Pinching-Antenna Systems,'' \textit{preprint}, 2025.

    \bibitem{Optimal_spacing} C. Ouyang, Z. Wang, Y. Liu, and Z. Ding, ``Array gain for pinchingantenna systems (PASS),'' \textit{arXiv preprint}, arXiv: 2501.05657, 2025.

    \bibitem{Discrete_matching} K. Wang, Z. Ding, and R. Schober, 
     ``Antenna Activation for NOMA Assisted Pinching-Antenna Systems,'' 
     \textit{arXiv preprint} arXiv:2412.13969, Dec. 2024.


    \bibitem{Channel_LoS} H. Elayan, O. Amin, B. Shihada, R. M. Shubair, and M. Alouini, 
    ``Terahertz band: The last piece of RF spectrum puzzle for communication systems,'' 
    \textit{IEEE Open J. Commun. Society}, vol. 1, pp. 1-32, Nov. 2020. 

    \bibitem{Spherical_Channel} H. Zhang, N. Shlezinger, F. Guidi, D. Dardari, M. F. Imani, and Y. C.
    Eldar, ``Beam focusing for near-field multiuser MIMO communications,''
    \textit{IEEE Trans. Wireless Commun.}, vol. 21, no. 9, pp. 7476-7490, Sept. 2022.

    \bibitem{Waveguide} 
    D. M. Pozar, \textit{Microwave engineering: theory and techniques}, John wiley \& sons, 2021. 

    \bibitem{WMMSE} Q. Shi, M. Razaviyayn, Z. -Q. Luo and C. He, 
    ``An Iteratively Weighted MMSE Approach to Distributed Sum-Utility Maximization for a MIMO Interfering Broadcast Channel,'' 
    \textit{IEEE Trans. Signal Process.}, vol. 59, no. 9, pp. 4331-4340, Sept. 2011.

    \bibitem{CCCP_NIPS}A. L. Yuille and A. Rangarajan, ``The concave-convex procedure (CCCP),'' in 
    \textit{Proc. Adv. Neural Inf. Process. Syst.}, 2002, vol. 2, pp. 1033-1040.

    \bibitem{MM_TSP}
    Y. Sun, P. Babu and D. P. Palomar, ``Majorization-Minimization Algorithms in Signal Processing, Communications, and Machine Learning,'' 
    \textit{IEEE Trans. Signal Process.}, vol. 65, no. 3, pp. 794-816, Feb. 1, 2017.
    
    \bibitem{Lipschitz_surrogate}
    Y. Xu and W. Yin, ``A block coordinate descent method for regularized multiconvex optimization with applications to nonnegative tensor factorization and completion,'' 
    \textit{SIAM J. Imaging Sci.}, vol. 6, no. 3, pp. 1758-1789, 2013.


    \bibitem{PDD}
    Q. Shi and M. Hong, 
    ``Penalty Dual Decomposition Method for Nonsmooth Nonconvex Optimization—Part I: Algorithms and Convergence Analysis,'' 
    \textit{IEEE Trans. Signal Process.}, vol. 68, pp. 4108-4122, 2020.

    \bibitem{Interior_Point}
    F. A. Potra and S. J. Wright, ``Interior-point methods,'' 
    \textit{J. Comput. Appl. Math.}, vol. 124, no. 1-2, pp. 281-302, 2000.

    \bibitem{L2O}
    T. Chen, X. Chen, W. Chen, H. Heaton, J. Liu, Z. Wang, and W. Yin, 
    ``Learning to Optimize: A Primer and A Benchmark,'' \textit{J. Mach. Learn. Res.}, vol. 23, no. 189, pp. 1-59, 2022.

    \bibitem{DeepUnfolding} A. Balatsoukas-Stimming and C. Studer, 
    ``Deep Unfolding for Communications Systems: A Survey and Some New Directions,'' 
    \textit{Proc. IEEE Int. Workshop Signal Process. Syst. (SiPS)}, Nanjing, China, Oct. 2019, pp. 266-271.

    \bibitem{DeepUnfoldingWMMSE} Q. Hu, Y. Cai, Q. Shi, K. Xu, G. Yu and Z. Ding, 
    `'`Iterative Algorithm Induced Deep-Unfolding Neural Networks: Precoding Design for Multiuser MIMO Systems,'' 
    \textit{IEEE Trans. Wireless Commun.}, 
    vol. 20, no. 2, pp. 1394-1410, Feb. 2021.

    \bibitem{ConvexOpt}
    S. Boyd and L. Vandenberghe, \textit{Convex Optimization}, Cambridge, U.K.: Cambridge Univ. Press, 2004.


    \bibitem{BF_Emil}
    E. Bj\"ornson, M. Bengtsson and B. Ottersten, ``Optimal Multiuser Transmit Beamforming: A Difficult Problem with a Simple Solution Structure [Lecture Notes],'' 
    \textit{IEEE Signal Process. Mag.}, vol. 31, no. 4, pp. 142-148, Jul. 2014.


    \bibitem{Trnasformer}
    A. Vaswani, N. Shazeer, N. Parmar, J. Uszkoreit, L. Jones, A. N. Gomez, L. Kaiser, and I. Polosukhin, 
    ``Attention is all you need,'' in \textit{Proc. Adv. Neural Inf. Process. Syst. (NeurIPS)}, 
    Long Beach, CA, USA, Dec. 2017, pp. 5998-6008.

    \bibitem{PASS_tutorial}
    Y. Liu, H. Jiang, X. Xu, Z. Wang, J. Guo, C. Ouyang, X. Mu, Z. Ding, A. Nallanathan, G. K. Karagiannidis, and R. Schober, 
    ``Pinching-Antenna Systems (PASS): A Tutorial,'' \textit{arXiv preprint arXiv:2508.07572}, Aug. 2025.

    \bibitem{SubConnectedHybridBF} X. Song, T. K{\"u}hne and G. Caire, 
    ``Fully-/Partially-Connected Hybrid Beamforming Architectures for mmWave MU-MIMO,'' 
    \textit{IEEE Trans. Wireless Commun.}, vol. 19, no. 3, pp. 1754-1769, Mar. 2020.

    \bibitem{PDD_HybridBF}
    Q. Shi and M. Hong, ``Spectral Efficiency Optimization for Millimeter Wave Multiuser MIMO Systems,''
    \textit{IEEE Trans. Signal Process.}, vol. 66, no. 11, pp. 2944-2958, June 2018.


    \bibitem{UAV_planning_RIS}
    J. Huang, B. Wu, Q. Duan, L. Dong and S. Yu, ``A Fast UAV Trajectory Planning Framework in RIS-assisted Communication Systems with Accelerated Learning via Multithreading and Federating,'' 
    \textit{IEEE Trans. Mobile Comput.}, vol. 24, no. 8, pp. 6870-6885, Aug. 2025.

    
    \bibitem{FederatedDRL}
    K. Guo, M. Wu, X. Li, Z. Lin and T. A. Tsiftsis, 
    ``Joint Trajectory and Beamforming Optimization for Federated DRL-Aided Space-Aerial-Terrestrial Relay Networks 
    With RIS and RSMA,'' \textit{IEEE Trans. Wireless Commun.}, vol. 23, no. 12, pp. 18456-18471, Dec. 2024.
    

    \bibitem{PrioritizedInformationBottleneck}
    Z. Fang, S. Hu, J. Wang, Y. Deng, X. Chen and Y. Fang, 
    ``Prioritized Information Bottleneck Theoretic Framework With Distributed Online Learning for Edge Video Analytics,'' 
    \textit{IEEE Trans. Netw.}, vol. 33, no. 3, pp. 1203-1219, June 2025.

    
\end{thebibliography}
\end{document}